\documentclass[longauth]{aa}

\usepackage{graphicx}
\usepackage{color}
\usepackage{float} 
\usepackage{comment} 
\usepackage{array} 
\usepackage{tabularx} 
\usepackage{lscape} 
\usepackage{geometry} 
\usepackage{booktabs} 
\usepackage{soul} 

\usepackage{txfonts}
\begin{document} 

   \title{The miniJPAS Survey: 
   The radial distribution of star formation rates in faint X-ray active galactic nuclei}


   \author{
          Nischal Acharya\inst{\ref{ins:dipc},\ref{ins:upvehu}}
          \and
          Silvia Bonoli\inst{\ref{ins:dipc},\ref{ins:ikerbasque}}
          \and
          Mara Salvato\inst{\ref{ins:mpe}}
          \and
          Ariana Cortesi\inst{\ref{ins:obrio},\ref{ins:ifrio}}
          \and
          Rosa M. González Delgado\inst{\ref{ins:iaa}}
          \and
          Ivan Ezequiel Lopez\inst{\ref{ins:difa},\ref{ins:inaf}}
          \and
          Isabel Marquez\inst{\ref{ins:iaa}}
          \and
          Ginés Martínez-Solaeche\inst{\ref{ins:iaa}}
          \and
          Abdurro{'}uf\inst{\ref{ins:johnhop},\ref{ins:stsci}}
          \and
          David Alexander\inst{\ref{ins:durham}}
          \and
          Marcella Brusa\inst{\ref{ins:difa},\ref{ins:inaf}}
          \and
          Jonás Chaves-Montero\inst{\ref{ins:ifae}}
          \and
          Juan Antonio Fernández Ontiveros\inst{\ref{ins:teruel}}
          \and
          Brivael Laloux\inst{\ref{ins:durham},\ref{ins:athens}}
          \and
          Andrea Lapi\inst{\ref{ins:sissa}}
          \and
          George Mountrichas\inst{\ref{ins:ifcanta}}
          \and
          Cristina Ramos Almeida\inst{\ref{ins:tenerife},\ref{ins:laguna}}
          \and
          Julio Esteban Rodríguez Martín\inst{\ref{ins:iaa}}
          \and
          Francesco Shankar\inst{\ref{ins:soton}}
          \and
          Roberto Soria\inst{\ref{ins:cass},\ref{ins:torino},\ref{ins:sydney}}
          \and
          José M. Vilchez\inst{\ref{ins:iaa}}
          \and
            Raul Abramo\inst{\ref{ins:fisicasaopaulo}}
            \and
            Jailson Alcaniz\inst{\ref{ins:ON}}
            \and
            Narciso Benitez\inst{\ref{ins:iaa}}
            \and
            Saulo Carneiro\inst{\ref{ins:fisicabahia}}
            \and
            Javier Cenarro\inst{\ref{ins:teruel}}
            \and
            David Cristóbal-Hornillos\inst{\ref{ins:teruel}}
            \and
            Renato Dupke\inst{\ref{ins:ON}}
            \and
            Alessandro Ederoclite\inst{\ref{ins:teruel}}
            \and
            A.\,Hernán-Caballero\inst{\ref{ins:teruel}}
            \and
            Carlos López-Sanjuan\inst{\ref{ins:teruel}}
            \and
            Antonio Marín-Franch\inst{\ref{ins:teruel}}
            \and
            Caludia Mendes~de~Oliveira\inst{\ref{ins:astrosaopaulo}}
            \and
            Mariano Moles\inst{\ref{ins:teruel}}
            \and
            Laerte Sodré Jr.\inst{\ref{ins:astrosaopaulo}}
            \and
            Keith Taylor\inst{\ref{ins:instruments4}}
            \and
            Jesús Varela\inst{\ref{ins:teruel}}
            \and
            Héctor Vázquez Ramió\inst{\ref{ins:teruel}}
       }

   \institute{Donostia International Physics Center (DIPC), Manuel Lardizabal Ibilbidea, 4, E-20018, San Sebastián, Spain \label{ins:dipc}\\
         \and
          University of the Basque Country UPV/EHU, Department of Theoretical Physics, Bilbao, E-48080, Spain \label{ins:upvehu}\\
         \and
           IKERBASQUE, Basque Foundation for Science, E-48013, Bilbao, Spain\label{ins:ikerbasque}\\
         \and
             Max-Planck-Institut für extraterrestrische Physik, Giessenbachstr. 1, 85748 Garching, Germany \label{ins:mpe}\\
         \and                                           
            Observatório do Valongo, Universidade Federal do Rio de Janeiro, 20080-090, Rio de Janeiro, RJ, Brazil \label{ins:obrio}\\     
        \and                                           
            Instituto de Fisica, Universidade Federal do Rio de Janeiro, 21941-972, Rio de Janeiro, RJ, Brazil \label{ins:ifrio}\\     
         \and 
            Instituto de Astrofísica de Andalucía (CSIC), P.O. Box 3004, 18080 Granada, Spain \label{ins:iaa}\\
         \and
            Instituto de Fisica de Cantabria (CSIC-Universidad de Cantabria), Avenida de los Castros, 39005 Santander, Spain \label{ins:ifcanta}\\
         \and                                           
            Dipartimento di Fisica e Astronomia "Augusto Righi", Università di Bologna, via Gobetti 93/2, 40129 Bologna, Italy \label{ins:difa}\\
         \and
            INAF - Osservatorio di Astrofisica e Scienza dello Spazio di Bologna, via Gobetti 93/3, 40129 Bologna, Italy \label{ins:inaf}\\
         \and
             Scuola Internazionale Superiore di Studi Avanzati, Via Bonomea 265, 34136 Trieste, Italy \label{ins:sissa}\\
         \and
            Centre for Extragalactic Astronomy, Department of Physics, Durham University, UK \label{ins:durham}\\
         \and
            Institute for Astronomy \& Astrophysics, National Observatory of Athens, V. Paulou \& I. Metaxa, 11532, Greece \label{ins:athens}\\
         \and
            School of Physics and Astronomy, University of Southampton, Highfield, Southampton SO17 1BJ, UK \label{ins:soton}\\
         \and
            Instituto de Astrofísica de Canarias, Calle Vía Láctea, s/n, E-38205, La Laguna, Tenerife, Spain \label{ins:tenerife}\\
         \and
            Departamento de Astrofísica, Universidad de La Laguna, E-38206 La Laguna, Tenerife, Spain \label{ins:laguna}\\
        \and
            Department of Physics and Astronomy, The Johns Hopkins University, 3400 N Charles St. Baltimore, MD 21218, USA \label{ins:johnhop}\\
        \and
            Space Telescope Science Institute (STScI), 3700 San Martin Drive, Baltimore, MD 21218, USA \label{ins:stsci}\\
        \and
            Institut de Física d’Altes Energies (IFAE), The Barcelona Institute of Science and Technology, 08193 Bellaterra (Barcelona), Spain \label{ins:ifae}\\
        \and
            Centro de Estudios de F\'isica del Cosmos de Arag\'on (CEFCA), Plaza San Juan 1, E--44001, Teruel, Spain \label{ins:teruel}\\
        \and
            College of Astronomy and Space Sciences, University of the Chinese Academy of Sciences, Beijing 100049, China \label{ins:cass}\\
        \and
            INAF-Osservatorio Astrofisico di Torino, Strada Osservatorio 20, I-10025 Pino Torinese, Italy \label{ins:torino}\\
        \and
            Sydney Institute for Astronomy, School of Physics A28, The University of Sydney, Sydney, NSW 2006, Australia \label{ins:sydney}\\
        \and
           Departamento de F\'isica Matem\'atica, Instituto de F\'{\i}sica, Universidade de S\~ao Paulo, Rua do Mat\~ao, 1371, CEP 05508-090, S\~ao Paulo, Brazil \label{ins:fisicasaopaulo}\\
        \and
           Observatório Nacional, Rua General José Cristino, 77, São Cristóvão, 20921-400, Rio de Janeiro, RJ, Brazil\label{ins:ON}\\
       \and
           Instituto de Física, Universidade Federal da Bahia, 40210-340, Salvador, BA, Brazil\label{ins:fisicabahia}\\
       \and
           Universidade de S\~ao Paulo, Instituto de Astronomia, Geofísica e Ciências Atmosféricas, Rua do Mat\~ao, 1226, 05508-090, S\~ao Paulo, SP, Brazil \label{ins:astrosaopaulo}\\
       \and
           Instruments4, 4121 Pembury Place, La Canada Flintridge, CA 91011, U.S.A\label{ins:instruments4}\\
}

   \date{Received 19 January, 2024; Accepted 06 May, 2024}

  \abstract
{We study the impact of black hole nuclear activity on both 
 the global and radial star formation rate (SFR) profiles in X-ray-selected active galactic nuclei (AGN) in the field of miniJPAS, the precursor of the much wider J-PAS project. Our sample includes 32 AGN with $z<0.3$ detected via the \textit{XMM-Newton} and \textit{Chandra} surveys. For comparison, we assembled a control sample of 71 star-forming (SF) galaxies with similar magnitudes, sizes, and redshifts.

To derive the global properties of both the AGN and the control SF sample, we used \texttt{CIGALE} to fit the spectral energy distributions derived from the 56 narrowband and 4 broadband filters from miniJPAS. We find that AGN tend to reside in more massive galaxies than their SF counterparts. After matching samples based on stellar mass and comparing their SFRs and specific SFRs (sSFRs), no significant differences appear. This suggests that the presence of AGN does not strongly influence overall star formation.

However, when we used miniJPAS as an integral field unit (IFU) to dissect galaxies along their position angle, a different picture emerges. We find that AGN tend to be more centrally concentrated in mass with respect to SF galaxies. Moreover, we find a suppression of the sSFR up to 1R$\mathrm{_e}$ and then an enhancement beyond 1R$\mathrm{_e}$, strongly contrasting with the decreasing radial profile of sSFRs in SF galaxies. This could point to an inside-out quenching of AGN host galaxies. 

Additionally, we examined how the radial profiles of the sSFRs in AGN and SF galaxies depend on galaxy morphology, by dividing our sample into disk-dominated (DD), pseudo-bulge (PB), and bulge-dominated (BD) systems. In DD systems, AGN exhibit a flat sSFR profile in the central regions and enhanced star formation beyond 1R$\mathrm{_e}$, contrasting with SF galaxies. In PB systems, SF galaxies show a decreasing sSFR profile, while AGN hosts exhibit an inside-out quenching scenario. In BD systems, both populations demonstrate consistent flat sSFR profiles.

These findings suggest that the reason we do not see differences on a global scale is because star formation is suppressed in the central regions and enhanced in the outer regions of AGN host galaxies. While limited in terms of sample size, this work highlights the potential of the upcoming J-PAS as a wide-field low-resolution IFU for thousands of nearby galaxies and AGN.
}


\maketitle
\section{Introduction}

    Active galactic nuclei (AGN), powered by the accretion of gas into their supermassive black holes (SMBHs) in the centers of galaxies, are among the brightest and most energetic objects in the Universe. Studies have shown that tight correlations exist between the SMBH mass and properties of the host galaxy \citep{magorrianDemographyMassiveDark1998,ferrareseFundamentalRelationSupermassive2000}. To understand the formation and fueling of AGN, it is therefore necessary to understand their immediate environments and, particularly, their host galaxies. Several works \citep{kauffmannUnifiedModelEvolution2000,hopkinsUnifiedMergerDrivenModel2006,merloniCOSMICEVOLUTIONSCALING2009} suggest that AGN-driven feedbacks are at the origin of such correlations. Hence, understanding the coevolution of SMBHs and their hosts is crucial to understanding when and how the galaxies formed and evolved. 
    
    Even though it has been established that SMBHs lie at the center of all massive galaxies \citep{kormendyCoevolutionNotSupermassive2013}, the processes that turn quiescent SMBHs into AGN are still being explored. {Major galaxy mergers have been touted as the most probable formation and fueling mechanism of AGN \citep{barnesDynamicsInteractingGalaxies1992,kauffmannUnifiedModelEvolution2000,springelFormationSpiralGalaxy2005,marulliModellingCosmologicalCoevolution2008,bonoliModellingCosmologicalCoevolution2009}. However, recent studies suggest that the triggering of AGN through mergers might depend on the redshift \citep{georgakakisHostGalaxyMorphologies2009, kocevskiCANDELSConstrainingAGNMerger2012, villforthMorphologiesAGNHost2014, marianMajorMergersAre2019}. Especially in the low redshift regime, the dependence of AGN on mergers seems to be minimal \citep{reichardLopsidednessPresentDayGalaxies2009,cisternasBULKBLACKHOLE2010, sabaterTriggeringOpticalAGN2015, wethersGalaxyMassAssembly2022}. Instead, minor mergers and secular processes could be behind the triggering of AGN in galaxies \citep{georgakakisHostGalaxyMorphologies2009, villforthHostGalaxiesLuminous2017, hernandez-toledoSDSSIVMaNGAIncidence2023}. Understanding the mechanisms that trigger AGN host galaxies is important for gaining insights into the regulation of star formation, morphology, and the chemical enrichment of galaxies.} 

    While there have been numerous investigations on the impact of AGN on the growth and star formation history (SFH) of galaxies, the conclusions of these studies are still under debate (see \citealt{alexanderWhatDrivesGrowth2012, kormendyCoevolutionNotSupermassive2013, heckmanCoEvolutionGalaxiesSupermassive2014}). Some studies show that AGN quench the star formation in host galaxies (e.g., \citealt{pageSuppressionStarFormation2012,bargerHostGalaxiesXray2015}), whereas others find enhanced star formation in AGN host galaxies (e.g., \citealt{mullaneyHIDDENAGNMAIN2012, kimEvidenceYoungStellar2019}). 
    \citet{kalfountzouObservationalEvidenceThat2017} and \citet{suhMultiwavelengthPropertiesType2019} meanwhile present evidence that AGN activity does not quench or suppress star formation, and instead stellar mass and jet power might be its main drivers. The impact of AGN on galaxies might also be due to the morphology of the galaxies they reside in. \citet{povicAGNhostGalaxyConnection2012} found that almost 50\% of AGN hosts are seen in massive spheroidal and bulge-dominated (BD) galaxies, and about 18\% reside in disk-dominated (DD) hosts  \citep{furusawaSubaruXMMNewtonDeep2008} using a sample of X-ray AGN hosts from the Subaru/\textit{XMM-Newton} Deep Survey (SXDS). \citet{rosarioXRAYSELECTEDAGN2013} found that, particularly in the X-ray AGN regime, the star formation rate (SFR) of X-ray-selected quasars is consistent with that of star-forming (SF) galaxies in the Cosmological Evolution Field Survey (COSMOS) field \citep{scovilleCosmicEvolutionSurvey2007}. \citet{masouraDisentanglingAGNStarFormation2018} meanwhile suggest that AGN enhance the star formation of their host galaxies when they lie below the main sequence of galaxies, and suppress star formation when the host galaxies lie above the main sequence for X-ray-selected AGN within 0.03 < z < 3. Recently, \cite{mountrichasComparisonStarFormation2022, mountrichasStarFormationXray2022} have found that X-ray AGN with 42 < $\mathrm{log}$ Lx$\mathrm{_{2-10keV}}$ < 44 tend to have SFRs that are lower than, or at most similar to, inactive galaxies, but tend to have higher SFRs than inactive galaxies at higher X-ray luminosities. The discrepancies arising from these studies could be a result of using different samples of AGN and comparison galaxies, or could indicate that the properties of host galaxies change as a function of redshift.
    
    These studies have provided context for the global coevolution of SMBHs and their host galaxies. To truly understand the impact of AGN on their host galaxies, however, it is necessary to also study effects at scales ranging from parsecs to kiloparsecs, especially in the central region where AGN activity can have the greatest impact \citep{ellisonEDGECALIFASurveyCentral2021, almeidaDiverseColdMolecular2022}. By resolving the galaxy into smaller spatial regions and analyzing the properties of these regions in detail, integral field spectroscopy (IFS) studies using integral field units (IFUs) can provide valuable insights into the feedback mechanisms of AGN. It has been shown that the distribution of the SFR in galaxies is not uniform but rather displays a complex spatial variation that depends on the environment and morphology of the galaxy \citep{delgadoCALIFASurveyHubble2015, delgadoSpatiallyresolvedStarFormation2017a}. Variations in the SFR of galaxies are usually associated with their gas kinematics, and metallicity can have a significant impact on the evolution of galaxies and the formation and properties of stellar populations and SF is regulated via various environmental processes, such as ram-pressure stripping and tidal interactions, and internal processes, such as feedback from AGN \citep{delgadoCALIFASurveyHubble2015,taylorMetallicityElementalAbundance2019,bessiereSpatiallyResolvedEvidence2022a}. These studies have shed new light on the complex interplay between the galaxies and their surrounding local environments, and have given valuable insights into the physical processes that regulate star formation and shape the properties of galaxies over cosmic time. As such, it is critical to investigate the impact of AGN on both the centers of their galaxies, where they reside, and the outskirts. Therefore, we expanded on these previous studies by examining the spatial impact of X-ray AGN on the galaxies they inhabit.
        
    In this study we examined the properties of galaxies from the central regions to their outskirts and explored the role of AGN activity in shaping their host galaxies. We studied the impact of AGN on the star formation of host galaxies, particularly at low redshifts, by obtaining their physical properties and comparing them to a carefully constructed control sample of inactive galaxies. We used the data provided by the miniJPAS survey in a square degree of area to construct a small sample of active and inactive galaxies selected by X-ray emission. We first compared their global properties and then investigated whether the behavior of radial profiles differ  in different X-ray luminosities and morphological systems. With this work, we also set the stage for a similar study with a much larger sample of active and inactive galaxies to be provided by the ongoing Javalambre Physics of the Accelerating Astrophysical Survey (J-PAS).
    
    The structure of this paper is as follows. The data used in this work are described in Sect. 2. The method and spectral energy distribution (SED) fitting processes are outlined in Sect. 3. The results are presented in Sect. 4, and we discuss the implications of these results and conclusions in Sect. 5 and 6, respectively. Throughout this work, we adopt the parameters of Lambda cold dark matter cosmology estimated by \citet{planckcollaborationPlanck2018Results2020}, with $h$ = 0.674, $\Omega_M$ = 0.315, and $\Omega_\Lambda$ = 0.685. Magnitudes are quoted in the AB system unless specified otherwise.

\section{Data and sample selection}

    In this section we describe the data used in our research. In Sect. \ref{sec:minijpas} we describe the miniJPAS survey and the optical data used to fit the SED of galaxies in this study. Section \ref{sec:xray} describes the X-ray data and the catalogs used to select AGN. Finally, we summarize and explain the process of our sample selection in Sect. \ref{sec:sampleselection}.

    \subsection{The miniJPAS and J-PAS surveys}
    \label{sec:minijpas}
    J-PAS \citep{benitezJPASJavalambrePhysicsAccelerated2014} is a wide-field cosmological survey using a 2.5-meter telescope and a 4.7 square degree camera with 1.2 gigapixels at the Javalambre Observatory in Spain. It is equipped with 54 narrowband filters, with a full-width half maximum (FWHM) of 145 {\AA}, complemented with two broadband filters in the blue and red optical wavelengths, that provide continuous spectral coverage from 3780 {\AA} to 9100 {\AA}. The survey will ultimately observe thousands of square degrees of the northern sky and measure photometric redshifts (photo-z) for more than 90 million galaxies and millions of AGN in an effective volume of approximately 14 Gpc$^3$ out to $z = 1.3$. 
    
\begin{figure}
        \includegraphics[width=0.5\textwidth]{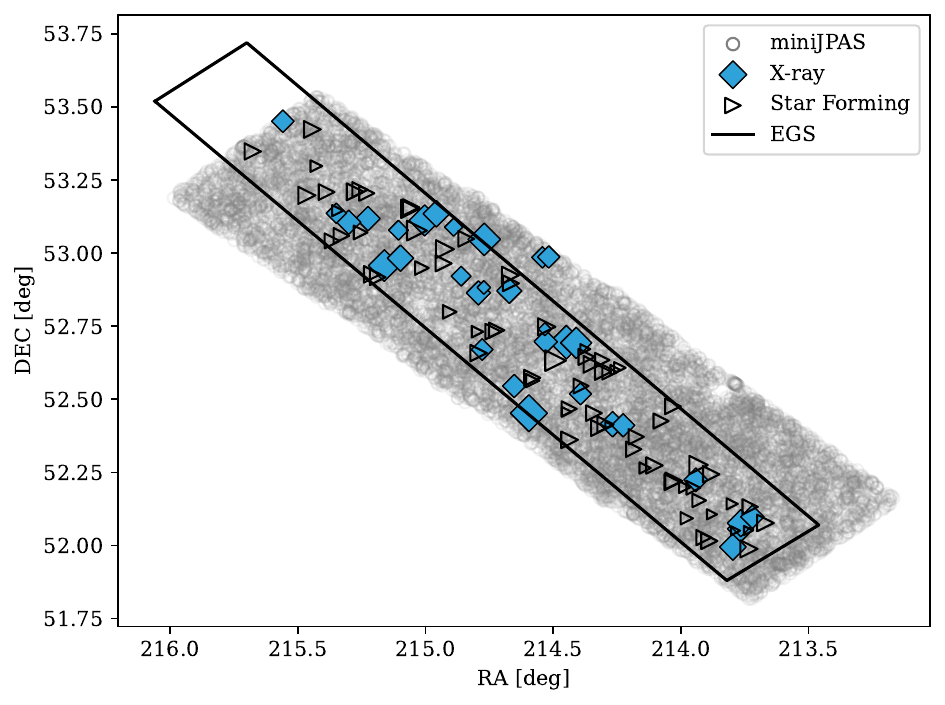}
        \caption{Sky map showing the location of the X-ray sample (blue diamonds) and the SF sample (empty triangles) in the miniJPAS footprint (gray). The rectangular area (black line) shows the EGS field \citep{davisAllWavelengthExtendedGroth2007}.}
        \label{fig:SkyMap}
    \end{figure}

    The miniJPAS survey \citep{bonoliMiniJPASSurveyPreview2021}, a preview of the full J-PAS survey, is a 3D survey covering 1 square degree of the {Extended Groth Strip (EGS)}/All-wavelength Extended Groth Strip International Survey (AEGIS) field \citep{davisAllWavelengthExtendedGroth2007} using 60 optical filters from J-PAS. Equipped with a 9k × 9k CCD and 0.3 square degree field of view and resolution of 0.23 arcseconds per pixel, it is complete to r = 23.6 AB for point-like sources and r = 22.7 AB for extended sources. The miniJPAS catalog includes more than 64,000 sources, with primary detection in the r band and forced photometry in all other bands.  Figure \ref{fig:SkyMap} shows the footprint of miniJPAS, the EGS field, and the positions of our galaxies hosting AGN and the control sample (see below). 

    {Studies {such as} \citet{romanJPLUS2DAnalysis2019a} and \citet{delgadoMiniJPASSurveyIdentification2021}  have demonstrated the capability of multiwavelength photometric surveys to study the global and spatial properties of stellar population of galaxies with the J-PLUS and miniJPAS datasets, respectively.}
    They have shown that an advantage of these datasets over the traditional IFUs is that, they provide an unbiased and wider field of view of galaxies that would otherwise not be observed by IFU surveys. The miniJPAS dataset has also already been used for studying several topics in the field of galaxy evolution such as their coevolution with SMBH \citep{chaves-monteroBlackHoleVirial2022,lopezMiniJPASSurveyAGN2023}, the role of environment in star formation \citep{delgadoMiniJPASSurveyRole2022} and selection of quasars using machine learning and artificial neural networks \citep{queirozMiniJPASSurveyQuasar2023, rodriguesMiniJPASSurveyQuasar2023, martinez-solaecheMiniJPASSurveyQuasar2023}. In this work, we utilized the IFS capabilities of miniJPAS to study the radial profiles of galaxies with and without an AGN.

    \subsection{Chandra and \textit{XMM-Newton}}
    \label{sec:xray}

    X-rays are excellent probes of AGN as they are not affected by the stellar light of host galaxies (see \citealt{brandtCosmicXraySurveys2015, padovaniActiveGalacticNuclei2017} for reviews).
    The EGS field has been well studied with X-ray missions, such as the X-ray Multi-Mirror Mission \citep[\textit{XMM-Newton};][]{liuSloanDigitalSky2020} and \textit{Chandra} surveys \citep{lairdAEGISXChandraDeep2009, nandraAEGISXDeepChandra2015} with a flux detection limit of $6.6 \times 10^{-16}$ erg/s/cm$^{2}$ and $5.5 \times 10^{-17}$ erg/s/cm$^{2}$ in 2-10 keV band, respectively \citep{ranalliXMMNewtonDeepSurvey2013}. Hence, we used the X-ray data from these two surveys to select AGN host galaxies in the miniJPAS field. We cross-matched the X-ray sources of the two surveys in both the soft (0.5-2 keV) and hard (2-10 keV) bands using a distance of five arcseconds, removing spectroscopically confirmed stars and spurious sources detected in the \textit{Chandra} survey following the \citet{nandraAEGISXDeepChandra2015} methods. The counterparts of \textit{Chandra} X-ray sources were identified using a likelihood estimation process with optical and infrared data in their study, while the counterparts of \textit{XMM-Newton} sources were identified using the Bayesian ``NWAY'' method \citep{salvatoFindingCounterpartsAllsky2018} as described in \citet{liuSloanDigitalSky2020}. In this work we used the X-ray fluxes of \textit{Chandra} sources when available as it has a higher spatial resolution of 0.4{"} FWHM compared to 6{"} FWHM of the XMM survey. A complete catalog of miniJPAS sources with X-ray emission detected in \textit{Chandra} and \textit{XMM-Newton} within $z = 2.5$ can be found in \citet{lopezMiniJPASSurveyAGN2023}.

    \subsection{Sample selection}
    \label{sec:sampleselection}
    
\begin{figure*}
        \includegraphics[width=1\textwidth]{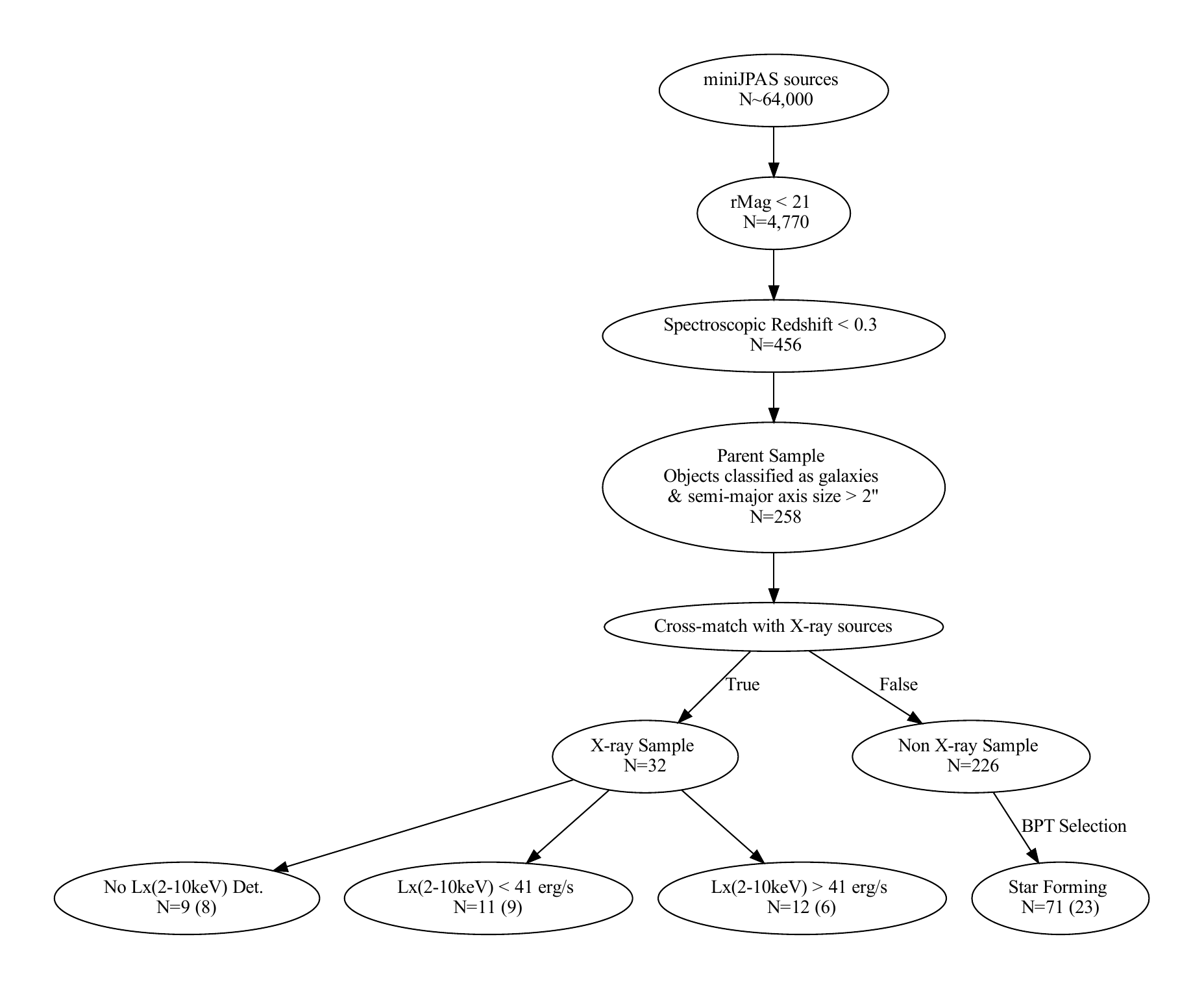}
        \caption{Steps taken to build the samples of active and inactive galaxies. No Lx$\mathrm{_{2-10keV}}$ detection signifies objects detected only in the 0.5-2keV band. The numbers in brackets denote the size of the mass-matched samples (see Sect. 4.1.1)}
        \label{fig:flowchart}
    \end{figure*}

\begin{figure*}
        \includegraphics[width=1\textwidth]{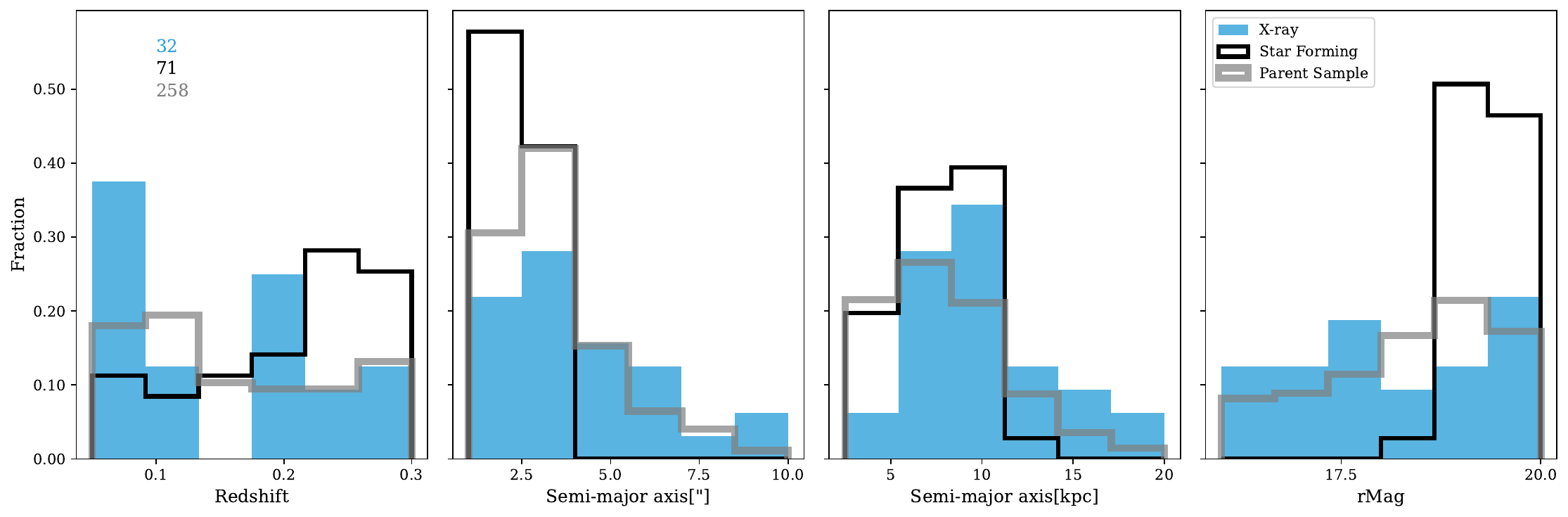}
        \caption{Distribution of AGN and SF samples. The redshift, size of the semimajor axis (in arcseconds and kiloparsecs), and the rMag of the  samples are shown. The gray histogram shows the distribution of the parent sample of miniJPAS galaxies.}
        \label{fig:SampleProperties}
    \end{figure*}

    The miniJPAS catalog contains 64,000 sources detected in the r-band. We needed sufficiently bright objects to be able to dissect galaxies elliptically to obtain their radial profiles, the goal of this work. Hence, we only selected objects with apparent r-band magnitude (rMag) < 21, which left us with 4,770 objects. We only used sources with spectroscopic redshift $z < 0.3$ in Sloan Digital Sky Survey (SDSS) Data Release 16 \citep{ahumadaSixteenthDataRelease2020} to eliminate any potential uncertainties on the physical properties of the objects. This leaves us with 456 objects with $z < 0.3$ in the local Universe. The magnitude and redshift cuts were chosen to maximize the quality of photometric data and maintain the visual quality of sources (S/N > 3), as we find most of the extended objects with high S/N lie within the aforementioned ranges in our sample. The miniJPAS catalog also provides us with the morphological classification, CLASS\_STAR values from SExtractor \citep{bertin1996} where extended sources are classified with CLASS\_STAR $\sim$ 0 and point sources with CLASS\_STAR $\sim$ 1. We selected extended sources with CLASS\_STAR < 0.1 and with semimajor size > 2" (min 0.82 kpc/" at $z=0.04$ to max 4.61 kpc/" at $z=0.3$ for our redshift range) as we needed extended objects with a big enough size to perform spatial operations. Finally, we were left with 258 galaxies within the limits of our selection process, which constitutes the parent sample.
    We present the sky map showing the locations of our X-ray and control samples within the miniJPAS footprint in Fig. \ref{fig:SkyMap}. The summary of the selection process of AGN and non-AGN samples can be seen in a flowchart in Fig. \ref{fig:flowchart}. Additionally, Fig. \ref{fig:SampleProperties} illustrates the distribution of redshift, semimajor axis size in both arcseconds and kiloparsecs and r-band magnitude distribution for both samples. 
    
    \subsubsection{X-ray sample}

\begin{figure}
        \includegraphics[width=0.5\textwidth]{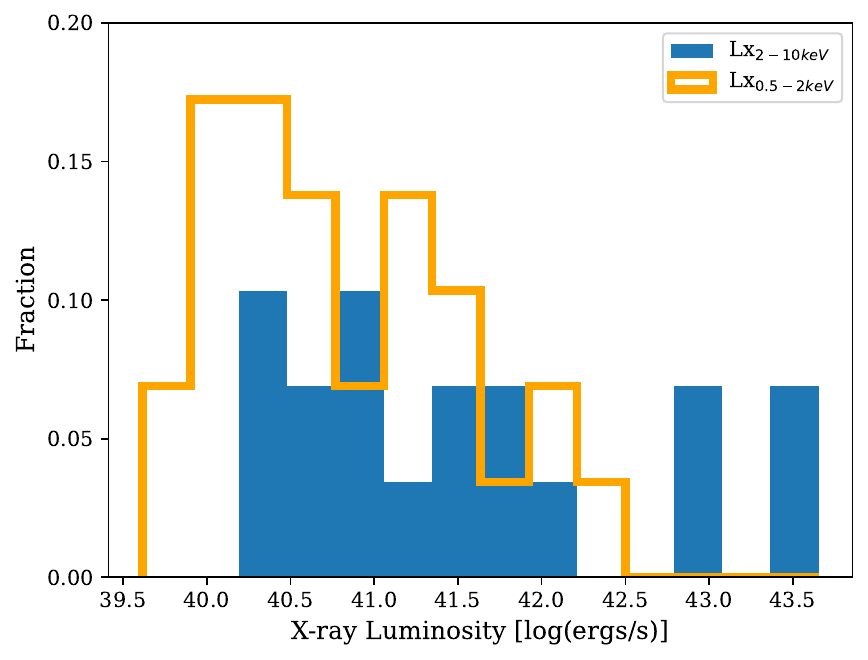}
        \caption{Distribution of X-ray luminosities in the soft (0.5-2 keV) and hard (2-10 keV) bands of the X-ray AGN sample. The plot shows that most of the selected AGN have weak X-ray emission, and only five objects have log Lx$\mathrm{_{2-10keV}}$ > 42 erg/s.}
        \label{fig:LxDistribution}
    \end{figure}

   To select AGN from the parent sample, we cross-matched them with the counterparts of X-ray sources from \textit{Chandra} and \textit{XMM-Newton} in the {0.5-2keV band within {5"}. This resulted in a total of 38 X-ray AGN within the defined magnitude, redshift, and size limits.} Upon visual inspection of the cross-matched sources, we found that six of them were corrupt in the miniJPAS database, which could be due to saturated pixels, overlapped tiles, or sources being too close to tile boundaries in the extraction process (see \citealt{bonoliMiniJPASSurveyPreview2021} for details on bad flags). Hence, the final AGN sample contains 32 X-ray detected galaxies. {A list of these galaxies is provided in Table \ref{tab:AGNsample}}. We further classified our AGN sample into three categories with respect to their X-ray luminosities. We find 9 objects without Lx$\mathrm{_{2-10keV}}$ X-ray detection in both surveys, 11 objects with log Lx$\mathrm{_{2-10keV}}$ < 41 erg/s and 12 objects with log Lx$\mathrm{_{2-10keV}}$ > 41 erg/s. The normalized distribution of X-ray luminosity of the AGN sources in soft and hard bands is shown in Fig. \ref{fig:LxDistribution}.

    \subsubsection{Control sample}

\begin{figure}
        \includegraphics[width=0.5\textwidth]{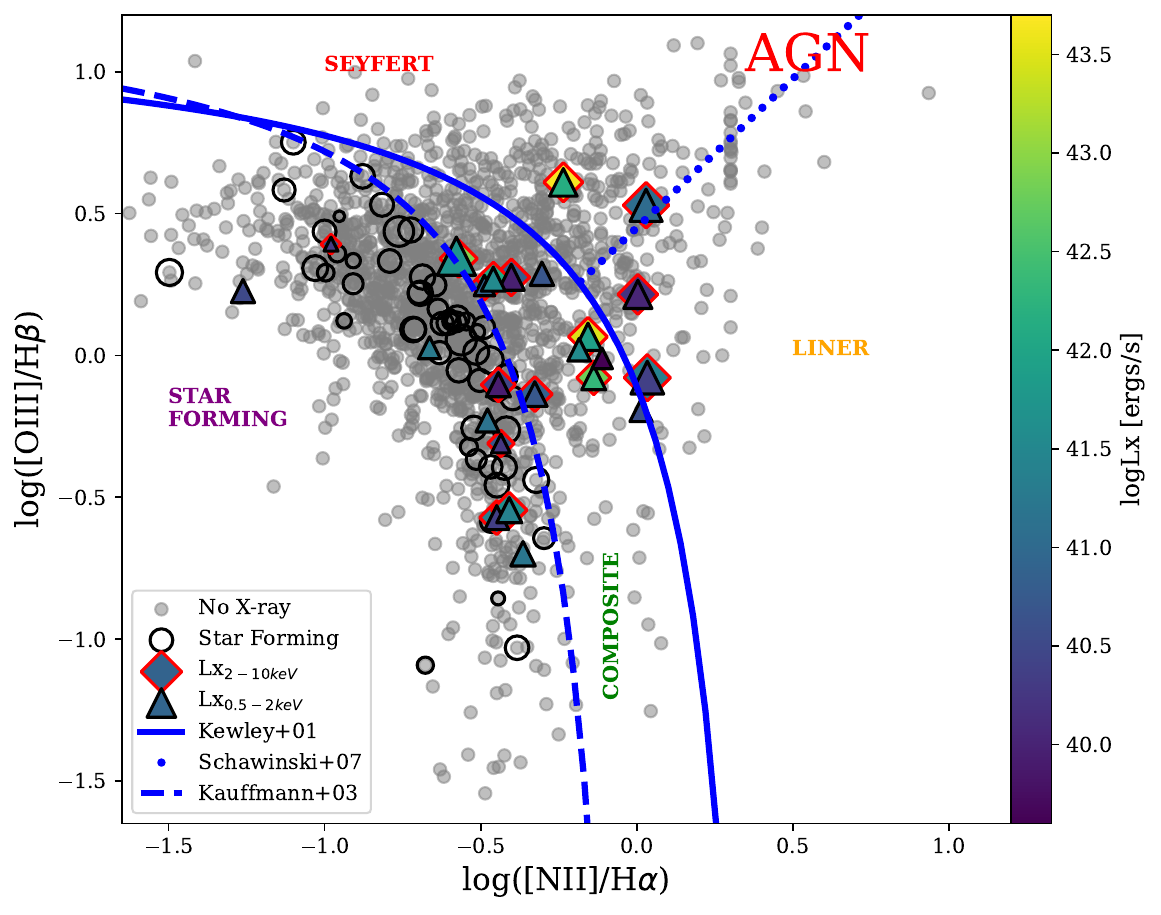}
        \caption{ BPT line diagnostic diagram of all the sources in our sample calculated from \cite{martinez-solaecheJPASMeasuringEmission2021}. The dashed blue line shows the SF line of Kauffman et al. (2003), and the solid blue lines show the maximum starburst model from Kewley et al. (2001). All the objects between these lines are known to be SF-AGN composites or transition objects. 
        The dotted line shows the demarcation line separating Seyferts and LINERs (Schawinski et al. 2007).
        The triangles show only soft (0.5-2 keV) X-ray-emitting galaxies, and the diamonds represent galaxies emitting both soft and hard (2-10 keV) X-ray fluxes. The gray points show all sources without X-ray emission, and the empty black circles below the dashed line show the SF sample. This figure also demonstrates the importance of selecting AGN using X-rays, as most of the AGN in our sample would have been identified as SF or composite galaxies if not for their X-ray detections.}
        \label{fig:BPT}
    \end{figure}

    The control sample allowed us to examine the impact of AGN activity on the host galaxies. On top of the X-ray-undetected galaxies {in the 0.5-2 keV band}, to ensure that our control sample consists of truly inactive galaxies, we excluded objects that are classified as optical AGN and composite galaxies based on the Baldwin, Phillips, and Terlevich (BPT) diagram \citep{baldwinClassificationParametersEmissionline1981}. The BPT diagram is a tool commonly used to differentiate between AGN and SF galaxies based on their emission line ratios. By comparing the [O III]/H$\beta$ and [N II]/H$\alpha$ line ratios of our sample, we could identify objects that show AGN-like emission and excluded them from the control sample. \citet{martinez-solaecheJPASMeasuringEmission2021} have developed a method based on Artificial Neural Network to measure the emission lines of J-PAS galaxies using synthetic photometry of Calar Alto Legacy Integral Field Area (CALIFA) \citep{sanchezCALIFACalarAlto2012}, Mapping Nearby Galaxies at Apache Point Observatory (MaNGA) \citep{bundyOverviewSDSSIVMaNGA2014}, and SDSS. The method has been tested and shown to identify and classify emission lines of galaxies in the miniJPAS field up to $z < 0.35$ in \citet{martinez-solaecheMiniJPASSurveyIdentification2022}. Excluding objects with X-ray counterparts, we were left with a total of 226 non X-ray galaxies. We used the \citet{kewleyTheoreticalModelingStarburst2001} and \citet{kauffmannHostGalaxiesAGN2003} lines as a boundary for classification. Sources lying between these lines are classified as composite samples, sources lying above the \citet{kewleyTheoreticalModelingStarburst2001} line are classified as optical AGN and sources falling below the \cite{kauffmannHostGalaxiesAGN2003} line are classified as SF samples. The SF sample represents the population of galaxies that have high star formation activity and negligible AGN activity, which we used as the control sample for this {work}, while the composite sample represents the population of galaxies that have both star formation and AGN activity. With this segregation, we were left with 71 purely SF galaxies. The distribution of all the extended sources in the BPT diagram used in this study can be seen in Fig. \ref{fig:BPT}.

\section{Methods}

    In this section we first describe the methods used in this work to model and subtract the contribution from the AGN to the galactic spectra using \texttt{GALFITM} in Sect. \ref{sec:removingagn}. Second, we describe the steps taken to dissect the galaxies into several elliptical annuli with piXedfit in Sect. \ref{sec:dissecting}. Finally, we describe the SED fitting procedure using the Code Investigating Galaxy Emission (\texttt{CIGALE}) in Sect. \ref{sec:SED}.


    \begin{figure*}[htbp]
        \includegraphics[width=1\textwidth]{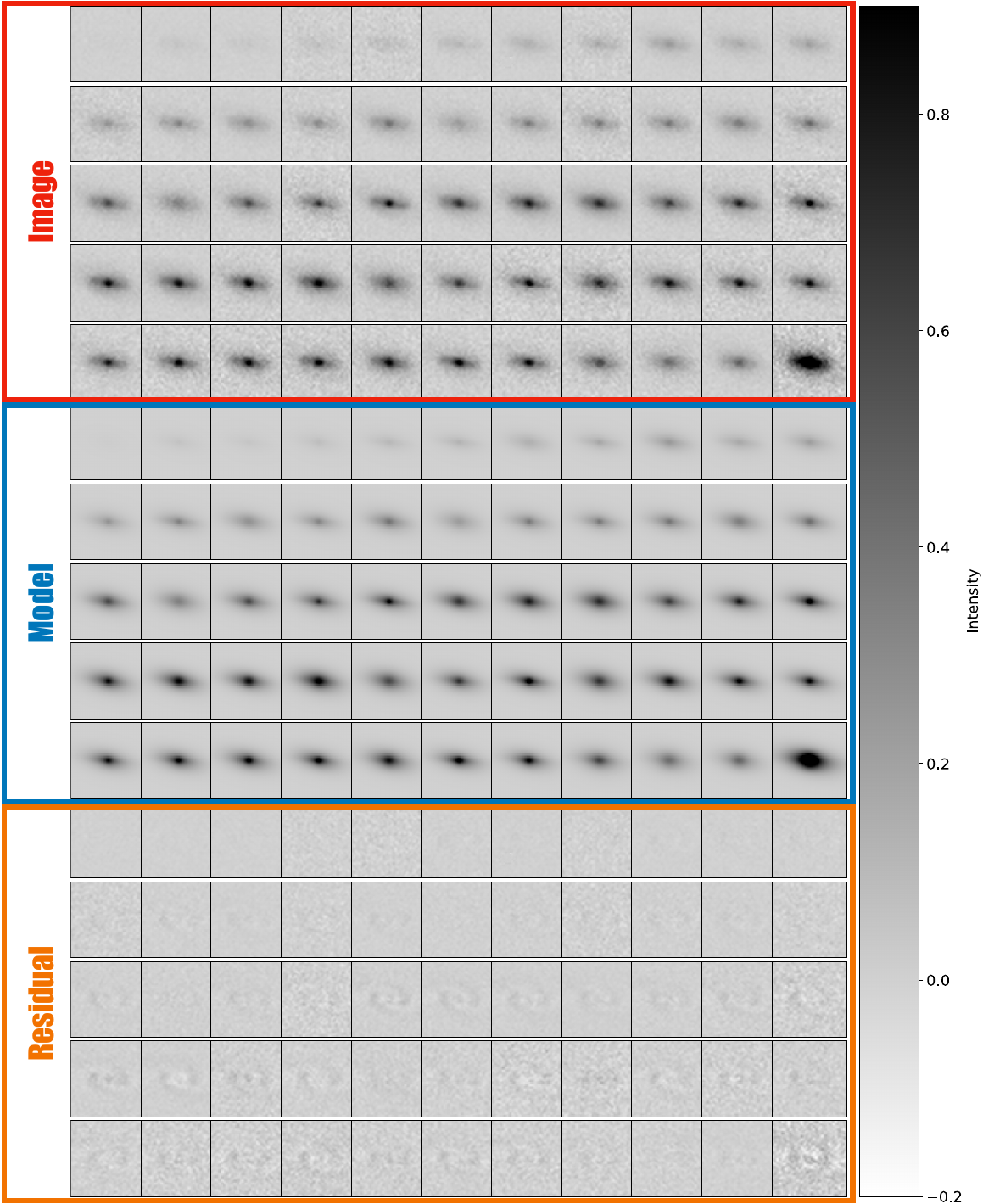}
        \caption{Example of multiwavelength three-component decomposition of an AGN host galaxy at z=0.11 (miniJPAS ID:2470-14395). The first five rows show the original images in the narrowband filters taken from the miniJPAS survey. The next five rows show the models generated from \texttt{GALFITM}, which were obtained by modeling the galaxy with a fixed galactic disk, a free Sérsic parameter for the bulge, and a point spread function  for the AGN component. The last five rows show the residual images when the model images in each filter are subtracted from the original images. }
        \label{fig:galfit}

    \end{figure*}

    \subsection{{Estimate of the AGN contribution}}
    
    \label{sec:removingagn}
    
    {AGN emission can dominate the light in the UV-optical wavelengths and mask the underlying properties of the host galaxies, especially in the case of type-I AGN \citep{suhMultiwavelengthPropertiesType2019}. Hence, it is essential to remove the contamination from the AGN emission to study the unbiased properties of the host galaxies of our X-ray AGN sample. To achieve this, we used \texttt{GALFITM} \citep{pengDetailedStructuralDecomposition2002,vika2014, hausslerGalapagos2GalfitmGAMA2022}, software that performs a multiwavelength morphological decomposition of galaxies by fitting a model to the observed image. This allowed us to separate the contribution of the AGN component from the rest of the galaxy.}

\begin{figure}
        \includegraphics[width=0.5\textwidth]{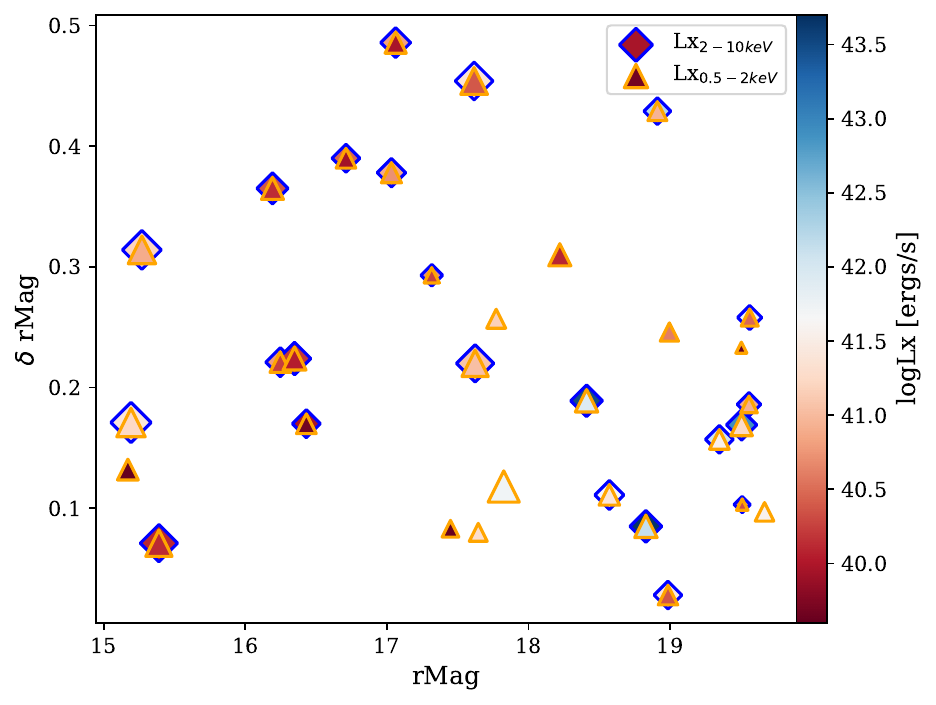}
        \caption{Scatter plot showing the difference in the r-band magnitude of AGN host galaxies before and after the removal of the {AGN contribution}. The points in the soft and hard bands are color-coded based on their X-ray luminosity. The size of the points is relative to the physical size of the host galaxies. The plot shows the impact of AGN in the observed magnitudes of galaxies. However, we do not see any correlation between the $\delta$ rMag and rMag or Lx of the AGN hosts.}
        \label{fig:XrayMagScatter}
    \end{figure}

    {We performed three-component modeling of the full galaxy, to make sure the AGN component is properly estimated, although ultimately we only subtracted the AGN component from each image.  
    We used a Sérsic function with Sérsic index $n=1$ to fit the galaxy disk, a Sérsic function with free Sérsic index for the bulge, and a point spread function for the AGN component. 
    Figure \ref{fig:galfit} shows an example of the \texttt{GALFITM} decomposition with these three components for one of the AGN sources. Figure \ref{fig:XrayMagScatter} shows the scatter plot of $\delta$rMag (i.e., the difference in rMag before and after the removal of the AGN component) against the observed rMag of the X-ray sources. However, probably due to the low luminosities of the AGN in our sample, we do not find any evidence of a correlation between $\delta$rMag and X-ray luminosity.}
    
    {We note that the residuals do not appear to be perfectly flat (i.e., the intensity of residuals $\neq$0)  in the images shown in Fig. \ref{fig:galfit}, and some structures that look like spiral arms of the galaxy become visible in the residuals. Moreover, it could  also be possible that we overestimated the central AGN component, which could lead to the underestimation of mass and SFR of the central bins.
    To address this issue,  we investigated the average residuals in the central bin for all the X-ray galaxies modeled through \texttt{GALFITM}. Figure \ref{fig:AverageResiduals} shows the intensity of residuals in the central bin in each of the optical bands. We find that the deviation of intensity from 0 is minimal, with an average of 0.015$\pm$0.048 units. Additionally, we tested omitting this step and find that doing so does not significantly change our results (see Figure \ref{fig:GalfitCheck}). This, again, is likely due to the low-luminosities of the AGN in our sample.} Obtaining perfect morphological fits for the structures of the AGN host galaxies, however, is beyond the scope of this work, as this exercise is only meant to reduce the contamination of the AGN component as much as possible for the subsequent SED fitting and for obtaining an estimate of the magnitude of the disk and bulge components in each galaxy, which allows us to classify them morphologically (see Sect. \ref{sec:morphology}).

    \subsection{Dissecting {galaxies}}
    \label{sec:dissecting}    
    The main goal of this work is to understand the spatial variation of star formation in X-ray AGN host galaxies. Thus, we needed to divide these {galaxies} into various regions to analyze them separately. To accomplish this, we used a method similar to that of IFUs and utilized the binning module in \texttt{piXedfit} \citep{abdurroufUnderstandingScatterSpatially2017, abdurroufIntroducingPiXedfitSpectral2021}, which is a Python package that provides tools for analyzing spatially resolved properties of galaxies using multiband imaging data alone or in combination with IFS data. 
    
\begin{figure*}
        \centering
        \includegraphics[width=0.8\textwidth]{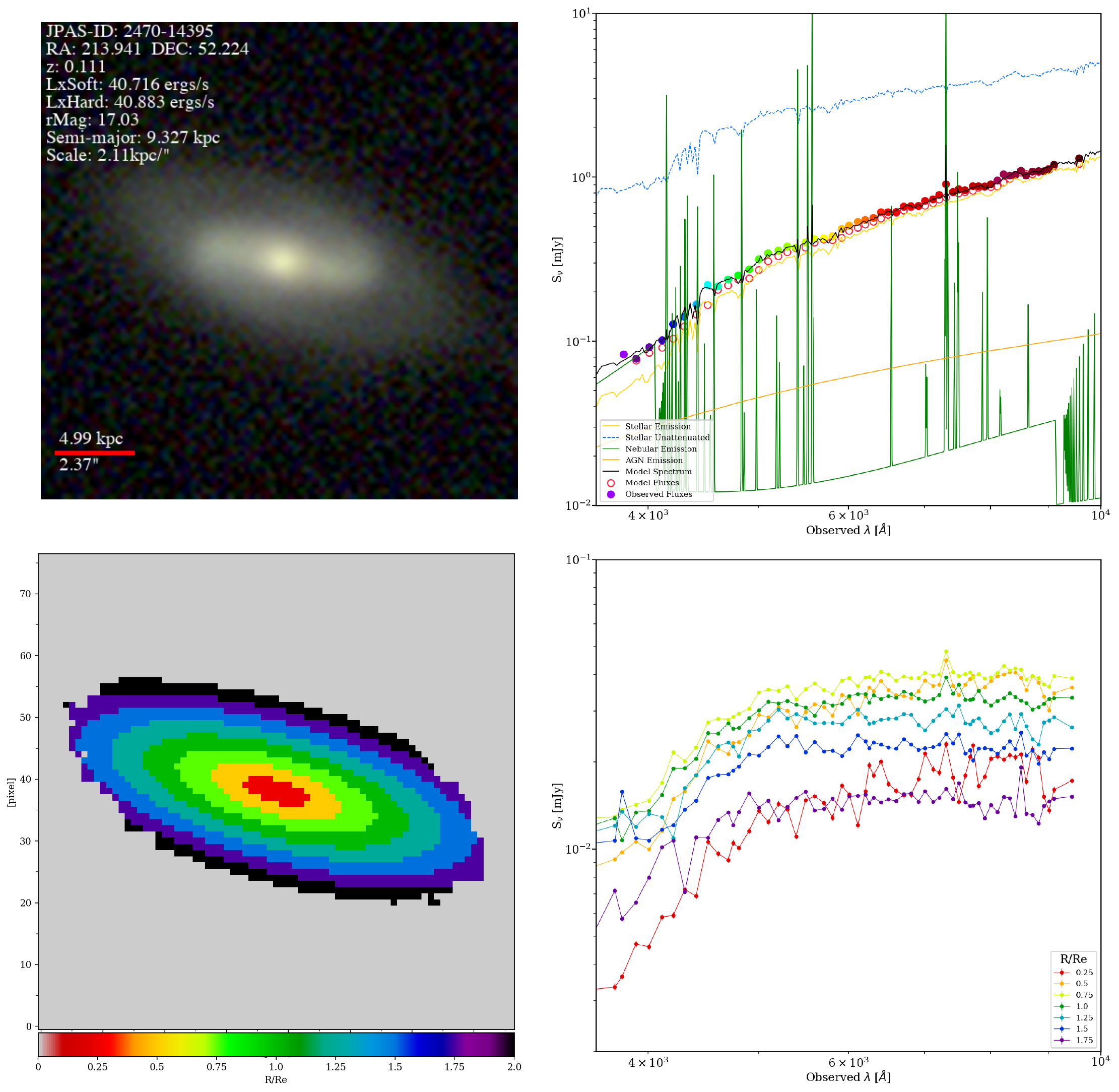}
        \caption{AGN host galaxy (miniJPAS ID:2470-14395)  at z = 0.11 (top left) and its J-spectra and SED fitting (top right). The filled colored points show the observed J-spectra, and the open red circles show the fitted points on the spectra. The host galaxy is divided into several elliptical annuli, as shown in the lower-left panel, and the resulting J-spectra for each of those shells are shown in the lower-right panel.}
        
        \label{fig:dissect}

    \end{figure*}

    {For each galaxy, we subtracted the AGN component (estimated by \texttt{GALFITM}) from the host galaxy and fed this image into \texttt{piXedfit}.
    Then, using the binning module in \texttt{piXedfit}}, we divided each of our sources into several radial bins based on their effective radii ({R$\mathrm{_e}$}) with respect to their position angle obtained from the output of \texttt{GALFITM} for r-band images, with each bin having a width of 0.25 R$\mathrm{_e}$ up to 1.75 {R$\mathrm{_e}$}. We limited our investigation for this work to 1.75 due to the decrease in the S/N as we move farther toward the edges of galaxies.  An example of this process is shown in Fig. \ref{fig:dissect} (third panel), which illustrates the division of one of our X-ray sources into various radial bins with increments of 0.25 {R$\mathrm{_e}$}. We then extracted the J-spectra for each of these bins and prepared them for SED fitting, as shown in the bottom right panel of Fig. \ref{fig:dissect}. The magnitude of each pixel is given by the equation 
    \begin{equation}
        magAB = -2.5log(ADU) + ZPT
    ,\end{equation}
    where ADU stands for analog-to-digital units and is used in the miniJPAS survey to store the values of each pixel in the survey area. The ADU keeps the record of the intensity of photons hitting the CCD of a telescope in a digital form and ZPT refers to the zero point of the filter in context. The total magnitude of each elliptical annulus can then be obtained by adding the contribution of each pixel in the respective bins. {The J-spectra obtained for each annulus is then fed to \texttt{CIGALE} for the SED fitting procedure to obtain the mass and SFR of the galaxies within the annuli.}
    
    \subsection{SED fitting}
    \label{sec:SED}
    Several programs and codes have been developed to facilitate the process of obtaining physical properties of galaxies. Most of these tools are based on the principle of energy balance, which states that the energy emitted in the infrared is equal to the energy absorbed in the ultraviolet/optical wavelengths, for example BAGPIPES \citep{carnallInferringStarformationHistories2018}, MAGPHYS \citep{dacunhaMAGPHYSPubliclyAvailable2011}, and ProSpect (\citealt{robothamProSpectGeneratingSpectral2020}). In this work we used \texttt{CIGALE} v2022 \citep{yangFittingAGNGalaxy2022}, an upgraded version of X-CIGALE \citep{BoquienCigale2019, yangXCIGALEFittingAGN2020}, which can take the luminosity produced by AGN into account in the SED fitting process. The results of SED fitting using X-CIGALE have been tested by \citet{mountrichasXrayFluxSED2021} for 2500 galaxies in the XMM-XXL field, and the models produced by X-CIGALE were found to agree with the results of visual inspection of randomly selected optical SDSS spectra for 85\% of AGN within z < 1 when using X-ray flux. The authors conclude that the inclusion of X-ray flux in the modeling improves the statistical significance of AGN fraction measurements. Even though the inclusion of X-ray significantly improves the estimation of stellar population for luminous AGN with Lx $>$ 10$^{45}$ erg s$^{-1}$, they found that including X-ray in SED modeling generally improves the estimation and characterization of the AGN component. Therefore, for the purposes of this work, we used \texttt{CIGALE} v2022 (hereafter referred to as \texttt{CIGALE}) to obtain the physical properties of our samples.
    
    \texttt{CIGALE} takes into account various physical processes that can contribute to the observed SED of a galaxy, such as SF, dust attenuation, and AGN activity. In the latest version of \texttt{CIGALE}, the code includes an improved treatment of the extinction of UV and optical emission in the polar regions of AGN and models the X-ray emission of galaxies. The process of SED modeling with \texttt{CIGALE} involves inputting a variety of parameters and assumptions about the galaxy, such as its SFH, metallicity, and initial mass function. \texttt{CIGALE} then uses these inputs to generate a model SED for the galaxy, which can be compared to the observed SED to infer the physical properties of the galaxy. In the following section, we describe the working principle of \texttt{CIGALE} and discuss specific inputs and assumptions that we used in this paper to model the SEDs of our active and inactive galaxies.

    \paragraph{\textbf{Components of SED fitting in \texttt{CIGALE}}\\\\}
    
    \texttt{CIGALE} generates a range of models for a given object at a given redshift, taking into account the input parameters provided by the user. To create these models, \texttt{CIGALE} first computes the SFH of the galaxy from a grid of values for SFR. It then uses this information to estimate the stellar spectrum and single stellar population models. To fit the spectrum, \texttt{CIGALE} estimates the nebular emission from the production of Lyman continuum photons and determines the stellar and nebular attenuation using an attenuation law. It calculates the luminosity absorbed by dust and applies the law of energy balance to estimate the dust emission in the mid and far infrared bands. If requested in the input parameters, \texttt{CIGALE} can also include the contribution of an active nucleus to the model.
    Finally, \texttt{CIGALE} uses the $\chi^2$ values as the maximum likelihood estimator to select the best-fit model for the given object, based on the prior assumptions made in the input parameters. \texttt{CIGALE} also gives two different outputs for the estimates of stellar parameters. It gives Bayesian values that incorporate the prior and uncertainties to obtain a probability distribution of possible values, and ``best'' values referring to the best-fitting value of the stellar parameters obtained from the SED fitting procedure. We used the Bayesian values outputted by \texttt{CIGALE} instead of the best values to account for the errors in the estimates of these parameters. The prior assumptions we used to compute the models for our samples of active and inactive galaxies are as follows: For the SFH of our objects, we adopted a delayed SFH model where the SFR is proportional to $t/\tau^{2}$ times the exponential of negative ${t/\tau}$, where $t$ is the difference between $t_{0}$ and the lookback time and $\tau$ is the exponentially folding time:
    \begin{equation}
        \text{SFR} \propto \frac{t}{\tau^2} \cdot \exp\left(-\frac{t}{\tau}\right)
    .\end{equation}
    
    This model has a peak SFR at ${t = \tau}$ and decreases smoothly thereafter. In addition, we also included an exponential burst model in our SFH modeling, following the approach of \citet{malekHELPModellingSpectral2018a} to represent the most recent period of star formation. The stellar emission was modeled using the \citet{bruzualStellarPopulationSynthesis2003} library and a \citet{salpeterLuminosityFunctionStellar1955} initial mass function. We considered three different levels of metallicity: 0.008, 0.02, and 0.05. The attenuated stellar emission was estimated using the \citet{calzettiDustContentOpacity2000} dust attenuation law, while the dust emission was modeled using the \citet{daleTwoparameterModelInfrared2014} templates. In order to model the AGN emission of our galaxies, we used the SKIRTOR template developed by \citet{stalevskiSKIRTORDatabaseModelled2012}. We considered both type 1 and type 2 AGN by allowing for viewing angles of 30 degrees and 70 degrees, respectively. We also applied the \citet{calzettiDustContentOpacity2000} law for the extinction of dust. To account for X-ray emission from the AGN host galaxies, we adopted a photon index of 1.8, as suggested by the X-ray source catalog compiled by \citet{lopezMiniJPASSurveyAGN2023}. \texttt{CIGALE} also accounts for the emission of X-ray from low- and high-mass X-ray binaries (LMXBs and HMXBs, respectively). Additionally, we set the max value of $\alpha_{ox}$, a measure of the X-ray to ultraviolet flux ratio, to 0.2, the maximum acceptable value according to the {${\alpha_{ox}}$-{L$_{2500\AA}$}} relation from \citet{risalitiCosmologyAGNCan2017}. This value is also used in previous studies by \citet{yangXCIGALEFittingAGN2020} and \citet{mountrichasRoleAGNObscuration2021, mountrichasComparisonStarFormation2022}. Beside the contamination in SED from an AGN that appears as a point source, AGN can also leave signs of an extended, scattered emission in the host galaxies, for example, in the form of Lyman-alpha lines in the surrounding nebular gas. {This can have a direct impact on the estimation of SFR of galaxies, especially in the high-redshift (z>2) regime where these lines are shifted to the optical wavelengths \citep{starkKECKSPECTROSCOPYFAINT2013, debarrosPropertiesLymanBreak2014}.} To account for this, \texttt{CIGALE} uses the nebular emission module estimated using the \citet{inoueRestframeUltraviolettoopticalSpectral2011} models, which take into account massive stars continuum and emission lines produced by high energy photons that ionize the surrounding gas. Table \ref{table:Cigale} lists the parameters used to model the SEDs for all our galaxies. We ran the SED fitting on our samples with and without including the AGN and X-ray modules in \texttt{CIGALE}, and did not find any significant changes in the results of our fitting (see Fig. \ref{fig:WithWithoutXrayMass}). These results are in agreement with \citet{mountrichasXrayFluxSED2021} who found no significant contributions of X-ray on the properties of AGN for galaxies with log Lx$\mathrm{_{2-10keV}}$ < 43 erg/s. To assure the quality of SED fittings and make sure that they do not affect the results of this study, we checked the reduced chi-square (${\chi^2}$) values of the fitted samples and found that all of the samples showed {${\chi^2}$} values of less than 5.5 (Fig. \ref{fig:chisquares}). Only 1 of the 32 objects in the X-ray sample and 11 of the 97 objects in the control sample have {${\chi^2}$} values greater than 3. We confirm that including the objects with {${\chi^2}$}>3 does not produce any differences in the final results of this work.

\begin{table*}
\centering
\small
\renewcommand{\arraystretch}{1.2}
\begin{tabularx}{1.0\textwidth}{Xp{3.5cm}p{4.5cm}}
\toprule
{Parameter} & {Model} & {Values} \\
\midrule
& {SFH: delayed model} & \\
\cmidrule{2-2}
E-folding time [Myr] & & 50, 100, 200, 500, 700, 1000, 3000 \\
Stellar age [Myr] & & 1000, 2000, 3000, 4000, 5000, 7000, 9000, 10000, 12000 \\
\midrule
& {SSP: Bruzual \& Charlot (2003)} & \\
\cmidrule{2-2}
Initial mass function & Salpeter (1995) & \\
Metallicity (Z) & & 0.008, 0.02, 0.05 \\
\midrule
& {Galactic dust extinction} & \\
\cmidrule{2-2}
Dust attenuation law & Calzetti et al. (2000) & \\
Reddening E(B-V) & & 0.0, 0.1, 0.2, 0.3, 0.4, 0.5, 0.6, 0.7, 0.8, 0.9 \\
\midrule
& {Galactic dust emission}: {Dale et al. (2014)} \\
\cmidrule{2-2}
Alpha slope in dMdust x U ${\alpha}$dU & & 1.0, 1.5, 2.0, 2.5, 3.0 \\
\midrule
& {Nebular}: {Inoue. (2011)} \\
\cmidrule{2-2}
log U & & -2.0\\
width lines & & 200 kms$^{-1}$\\
\midrule
& {AGN module: Skirtor} & \\
\cmidrule{2-2}
Torus optical depth at 9.7 microns & & 7.0 \\
Torus density radial parameter (p) & & 1.0 \\
Torus density angular parameter (q) & & 1.0 \\
Angle between the equatorial plane and edge of the torus & & 40\textdegree \\
Ratio of the maximum to minimum radii of the torus & & 20 \\
Viewing angle & & 30\textdegree, 70\textdegree \\
AGN fraction & & 0.0, 0.01, 0.1, 0.2, 0.3, 0.4, 0.5, 0.6, 0.7, 0.8, 0.9, 0.99 \\
Extinction law of polar dust & SMC & \\
E(B-V) of polar dust & & 0.0, 0.01, 0.02, 0.03, 0.05, 0.1, 0.2, 0.4, 0.6, 1.0, 1.8 \\
Temperature of polar dust (K) & & 100 \\
Emissivity of polar dust & & 1.6 \\
\midrule
& {X-ray module} & \\
\cmidrule{2-2}
AGN photon index & & 1.8 \\
Maximum deviation from the L2500 relation & & 0.2 \\
LMXB photon index & & 1.56 \\
HMXB photon index & & 2.0 \\
\bottomrule
\end{tabularx}
\caption{\texttt{CIGALE} input parameters used in the SED modeling of galaxies.}
\label{table:Cigale}
\end{table*}

\section{Results}

    In Sect. \ref{sec:results_global} we report the results of the SED fitting for the total photometry of the galaxies, where we determine the stellar masses and SFR for each of our sample based on the flux measured inside the Kron radius from the miniJPAS catalog. Following the elliptical binning procedure outlined in Sect. \ref{sec:dissecting}, we display the radial profiles of these objects in Sect. \ref{sec:results_radial}.

    \subsection{Global properties}
    \label{sec:results_global}
    Here we present the results of our analysis of the stellar masses, SFRs, and specific star formation rates (sSFRs) of the AGN and SF populations. These quantities were obtained by performing SED fitting on the data, as described in Sect. \ref{sec:SED}.
    
    \begin{figure}
        \includegraphics[width=0.5\textwidth]{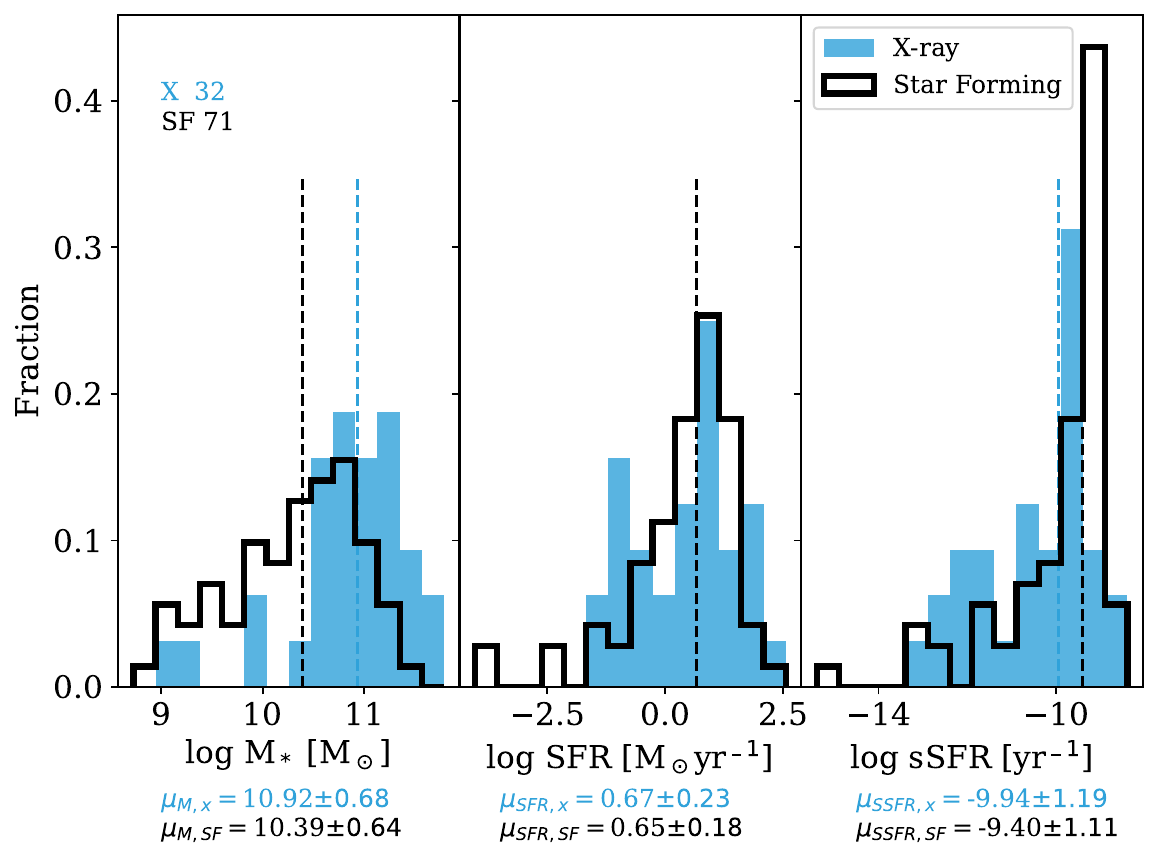}
        \caption{Distribution of the estimated stellar mass (left), SFR (center), and sSFR (right) for the X-ray AGN (blue), and SF (black) populations. The median values for each population are indicated by the vertical lines, and the corresponding values along with their associated errors are shown on the plot.}
        \label{fig:GlobalResults}
    \end{figure}

    Figure \ref{fig:GlobalResults} (left) shows the mass distribution of the X-ray AGN host galaxies compared to that of the SF sample. It is clear from the figure that the AGN tend to reside in more massive galaxies with a median of 10.92 $\pm$ 0.68 log M$_\odot$ than the SF sample with a median mass value of 10.39 $\pm$ 0.64 log M$_\odot$. KS test reveals a $p$ value of $\sim$10$^{-4}$, which shows that the two samples are not drawn from the same underlying population.
     The results are rather different when it comes to the star formation properties of these populations. From the central panel on Fig. \ref{fig:GlobalResults}, it appears that within 1 sigma, the median SFR for AGN and SF sample are the same. AGN sample shows a higher log SFR median values of 0.67 $\pm$ 0.23 compared to 0.65 $\pm$ 0.18 of the SF sample. However, it can be seen that the SFR distribution of the X-ray sample exhibits a bimodal distribution. Hence, any averages taken populations needs to be treated with caution.
    
    To get a more comprehensive view of the SF properties of these galaxies, we plotted the distribution of the sSFRs. The sSFR is defined as the SFR per unit mass ($\text{sSFR} = {\text{SFR}}/{\text{Mass}}$) and is often used to distinguish between actively SF and quiescent galaxies. sSFR is an important property of galaxies that is related to a characteristic time, is independent of the cosmology and initial mass function, and represents the time needed to grow a unit of stellar mass at the current SFR \citep{delgadoCALIFASurveyHubble2015, delgadoSpatiallyresolvedStarFormation2017a} . The distribution of the sSFRs for the AGN and SF samples, shown in Fig. \ref{fig:GlobalResults} (right), reveals a small difference between the two populations. The AGN population appears to have a larger proportion of sources with lower sSFRs, leading to a slightly smaller median value than that of the SF sample, with the AGN sample having a log median sSFR of $\sim$ -9.94 $\pm$ 1.19 yr$^{-1}$ compared to -9.40 $\pm$ 1.11 yr$^{-1}$ for the SF sample. The sSFR distribution of X-ray AGN shows two peaks at $\sim$ -11.9 yr$^{-1}$ and $\sim$ -9.6 yr$^{-1}$, which hints at two subpopulations in this sample. The presence of a larger fraction of quiescent galaxies in the AGN sample compared to the SF sample is clearly indicated by these results.

\begin{figure}
            \includegraphics[width=0.5\textwidth]{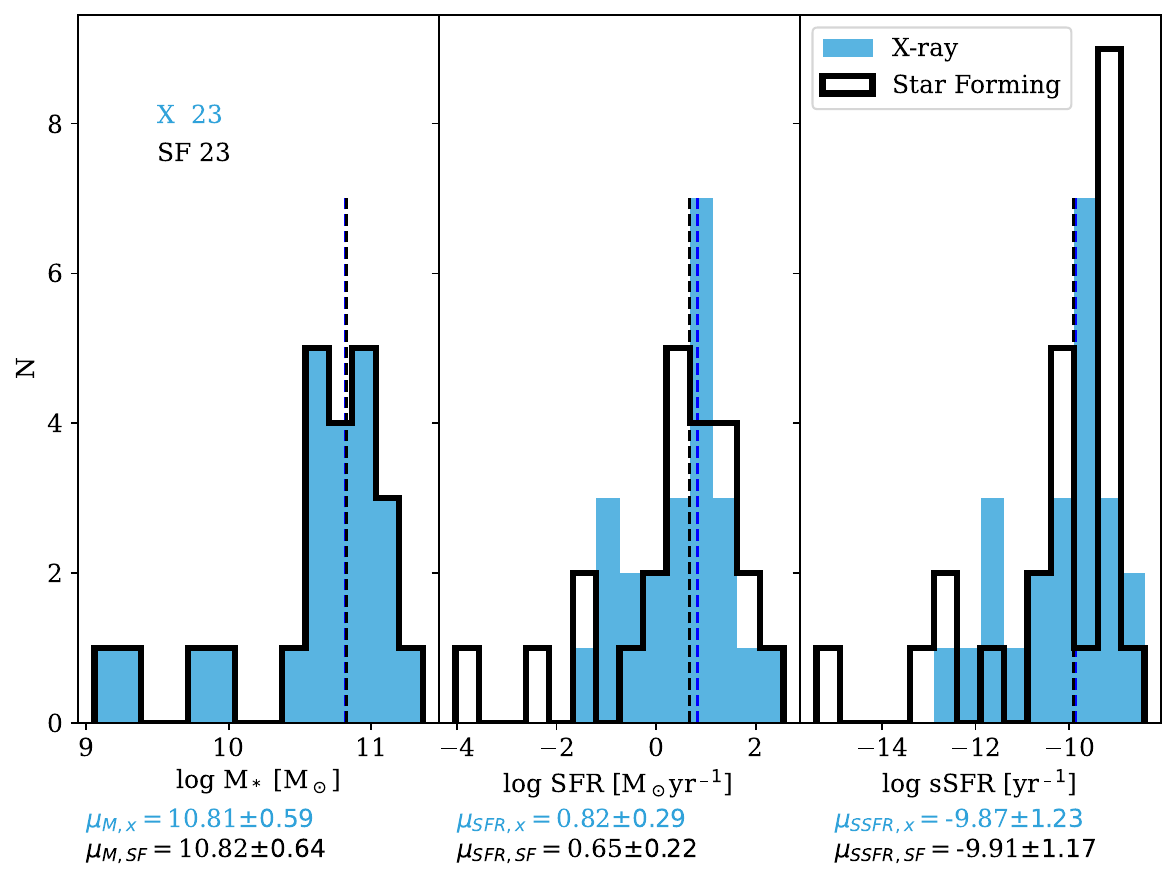}
            \caption{Distribution of the estimated stellar mass (left), SFR (center), and sSFR (right) for X-ray AGN (blue) and SF (black) mass-matched populations. The median values for each population are indicated by the vertical lines, and the corresponding values along with their associated errors are shown on the plot.}
            \label{fig:MassMatchedResults}
        \end{figure}

   \paragraph{\textbf{Effect of mass on SF parameters\\\\}}
    We find that X-ray AGN host galaxies have a preference for residing in more massive galaxies compared to SF galaxies. We can see in Fig. \ref{fig:GlobalResults} that the median mass of AGN hosts is $\sim$ 0.5 dex larger than the SF population, which most likely is due to our selection bias {arising from the apparently brighter AGN sample (see the rightmost panel of Fig. \ref{fig:SampleProperties})}. This might affect the results we obtain for the radial profiles of the samples, as massive galaxies are known to have higher SFR rates in general. In order to account for this difference, we selected SF galaxies whose masses were within a 0.2 dex range of AGN masses in each mass bin. For each of the mass bins, a random SF galaxy was chosen until the number of SF galaxies and AGN hosts were the same in those bins. Mass bins where we did not find enough SF samples to match against the AGN sample were discarded. The obtained distribution for global stellar mass, SFR and sSFR can be seen in Fig. \ref{fig:MassMatchedResults}. 

    We find that the global results change significantly when the samples are mass-matched ($p{\approx}1$). The AGN sample shows 0.2 dex higher median SFR compared to the SF sample, but KS test $p$ value of 0.99 indicates that they are drawn from the same underlying population. The difference in their sSFR is $\sim$0.04 dex, which is well within their error ranges of $\sim$1 dex ($p{\approx}0.42$). It is evident from these plots (Fig. \ref{fig:MassMatchedResults}) and KS tests that the SF and AGN samples with similar masses exhibit very similar SFRs, and AGN activity does not seem to have any effect on the SFH of their host galaxies on a global scale. We find this result is in line with findings reported by \citet{mountrichasComparisonStarFormation2022,mountrichasStarFormationXray2022} in their study of the influence of AGN on host galaxies, where they also did not find any significant differences in the SF properties of X-ray-selected AGN with 42 < log Lx$\mathrm{_{2-10keV}}$ < 44 and mass-matched SF galaxies. We argue that our results extend this finding to lower luminosities down to log Lx$\mathrm{_{2-10keV}} {\sim}$ 40 erg/s.
    However, some follow-up studies with much larger samples in the J-PAS field are required to confirm the results as the sample size used in this study is very small compared to the sample used in the \citet{mountrichasComparisonStarFormation2022,mountrichasStarFormationXray2022} papers ($\sim$1800 and 1000 galaxies, respectively).

    \subsection{Radial or 2D properties}
    \label{sec:results_radial}
    
    In this section we present the analysis of the radial properties of SF and AGN host galaxies. To ensure a fair comparison, we compared only the properties of the mass-matched samples presented in Sect. 4.1. {The average {R$\mathrm{_e}$} of the AGN sample was found to be 3.9$\pm$0.4 kpc compared to the average {R$\mathrm{_e}$} of 4.2$\pm$ 0.6 kpc of the SF galaxies.} We took any potential biases that could arise from the difference in their size into account  by normalizing the masses and SF properties of all the samples with respect to their surface area within the annuli, also known as their surface densities ($\Sigma$). We compared the radial variation in the $\Sigma_{M_\star}$, $\Sigma_{SFR}$, and sSFR of the AGN sample against those of the control sample to explore if they exhibit differences in their radial profiles. While sSFR is independent of cosmology, $\Sigma_{SFR}$ is largely insensitive to the past SFH and provides a measure of current SF activities and is tied to the molecular gas densities \citep{salimSigma_MathrmSFR2023a}. The combination of sSFR and $\Sigma_{SFR}$ provides a more comprehensive understanding of the SF activities of a galaxy.

\begin{figure}
        \includegraphics[width=0.5\textwidth]{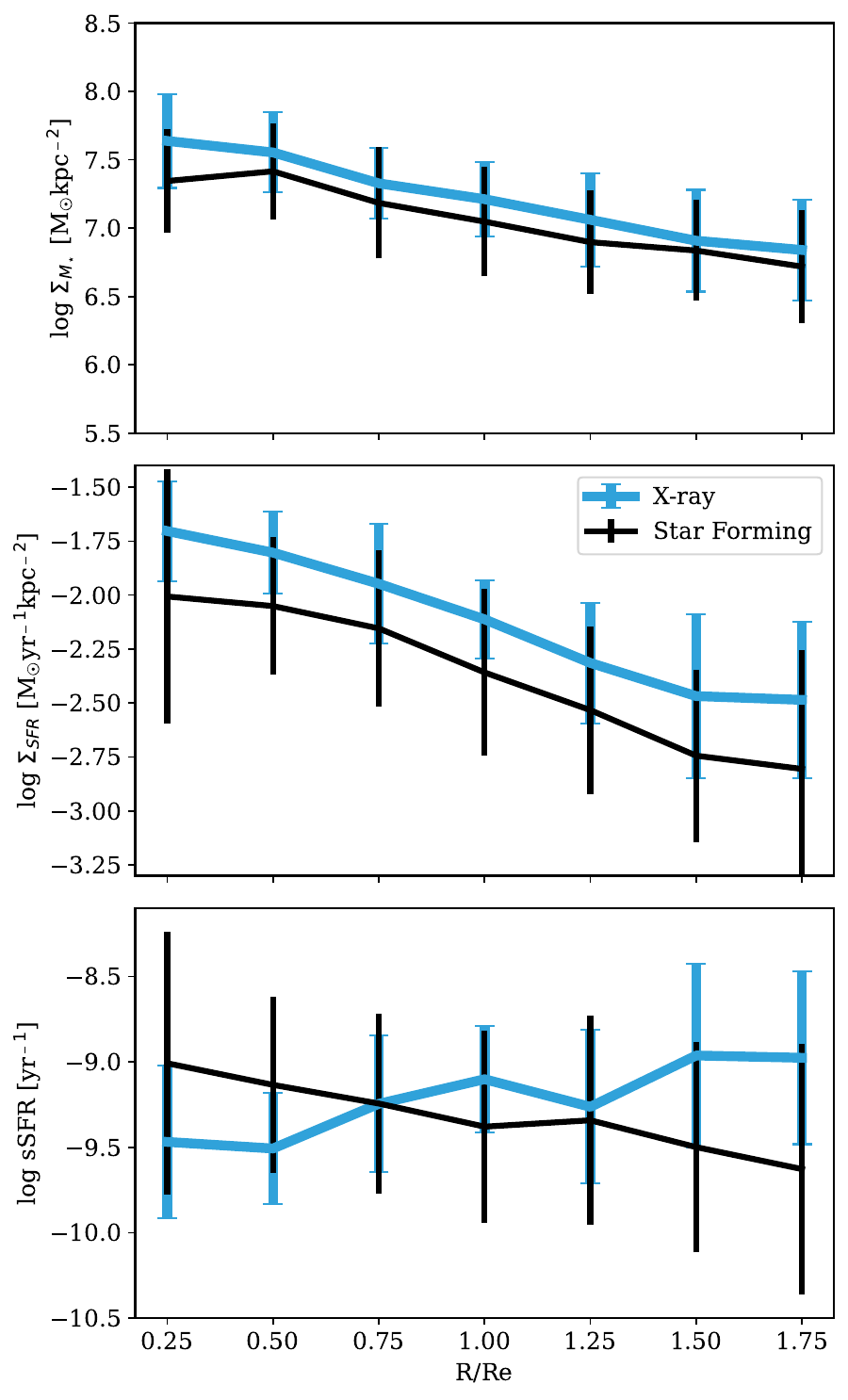}
        \caption{Radial profiles of the $\Sigma_{M_\star}$ (top), $\Sigma_{SFR}$ (middle), and sSFR (bottom) of AGN and the mass-matched SF sample. We can see that the mass profiles show similar characteristics. In other words, we can see that the AGN are still slightly more compact than the SF sample, but consistent within the errors. AGN display a higher $\Sigma_{SFR}$ profile compared to that of the SF sample. Their sSFR profiles seem to suggest that X-ray AGN enhance star formation in the outer regions of their host galaxies and quench the center, whereas we find the exact opposite scenario for the SF sample. Error bars represent the 3$\sigma$ confidence intervals.}
        \label{fig:MassMatchProfile}
    \end{figure}

    Figure \ref{fig:MassMatchProfile} shows the radial profiles for the samples matched by their stellar masses. We find that the X-ray-selected AGN host galaxies live in denser galaxies compared to the SF sample. They seem to be denser by $\sim$ 0.1 dex (but consistent within 3$\sigma$) than the SF population in the central regions and appear to diffuse toward the outskirts of galaxies (see Fig. \ref{fig:MassMatchProfile}, top panel). We find that the differences in their $\Sigma_{SFR}$ profiles also follow a similar trend, albeit with the X-ray population exhibiting slightly higher values throughout the annuli (Fig. \ref{fig:MassMatchProfile}, second panel). 
    
    The bottom panel in Fig. \ref{fig:MassMatchProfile} shows the sSFR profiles of the AGN and SF samples. Despite the fact that they seem to be consistent within 3$\sigma$, we observe that they grow in the exact opposite trend radially with respect to each other. Figure \ref{fig:MassMatchProfile} (bottom panel) shows that star formation is suppressed in the central regions and enhanced in the outer regions, contrasting with the sSFR profile of SF galaxies (i.e., AGN quench their host galaxies from the inside out). This is a very interesting finding and it suggests that X-ray AGN enhance star formation in host galaxies in the outskirts, and quench star formation in the central regions. Our results support recent works like \citet{ellisonEDGECALIFASurveyCentral2021} who have also found that the central kiloparsec-scale AGN regions have lower gas fractions than the outer SF regions. They argue that we do not see a difference in the global SF properties of active and inactive galaxies as the non-AGN regions in an otherwise active galaxy greatly overshadow the central AGN regions. Similar sSFR radial profiles were also found by \citet{jinIFUViewActive2021} in the MaNGA survey, which supports the idea of inside-out quenching in AGN host galaxies. Studies such as \citet{almeidaDiverseColdMolecular2022} and \citet{speranzaWarmMolecularIonized2022} calculated the rate and extent of molecular outflows in quasars and claim the outflows are AGN-driven, which would also explain the suppression of SF in central regions of host galaxies. Figure \ref{fig:MassMatchProfile} (bottom panel) suggests that the reason we do not see a difference in the global SF properties of active and inactive galaxies is due to the simultaneous balancing of suppression and enhancement of SF in AGN hosts. We also report that the profiles of SF galaxies are in agreement with \citet{delgadoStarFormationHubble2016} where they found a decreasing $\Sigma_{SFR}$ profile for SF galaxies in the CALIFA survey and showed that normal SF galaxies gradually slow down their SF processes toward the edges of galaxies.
    
    \subsubsection{Strong and weak X-ray galaxies}
    
\begin{figure}
        \includegraphics[width=0.5\textwidth]{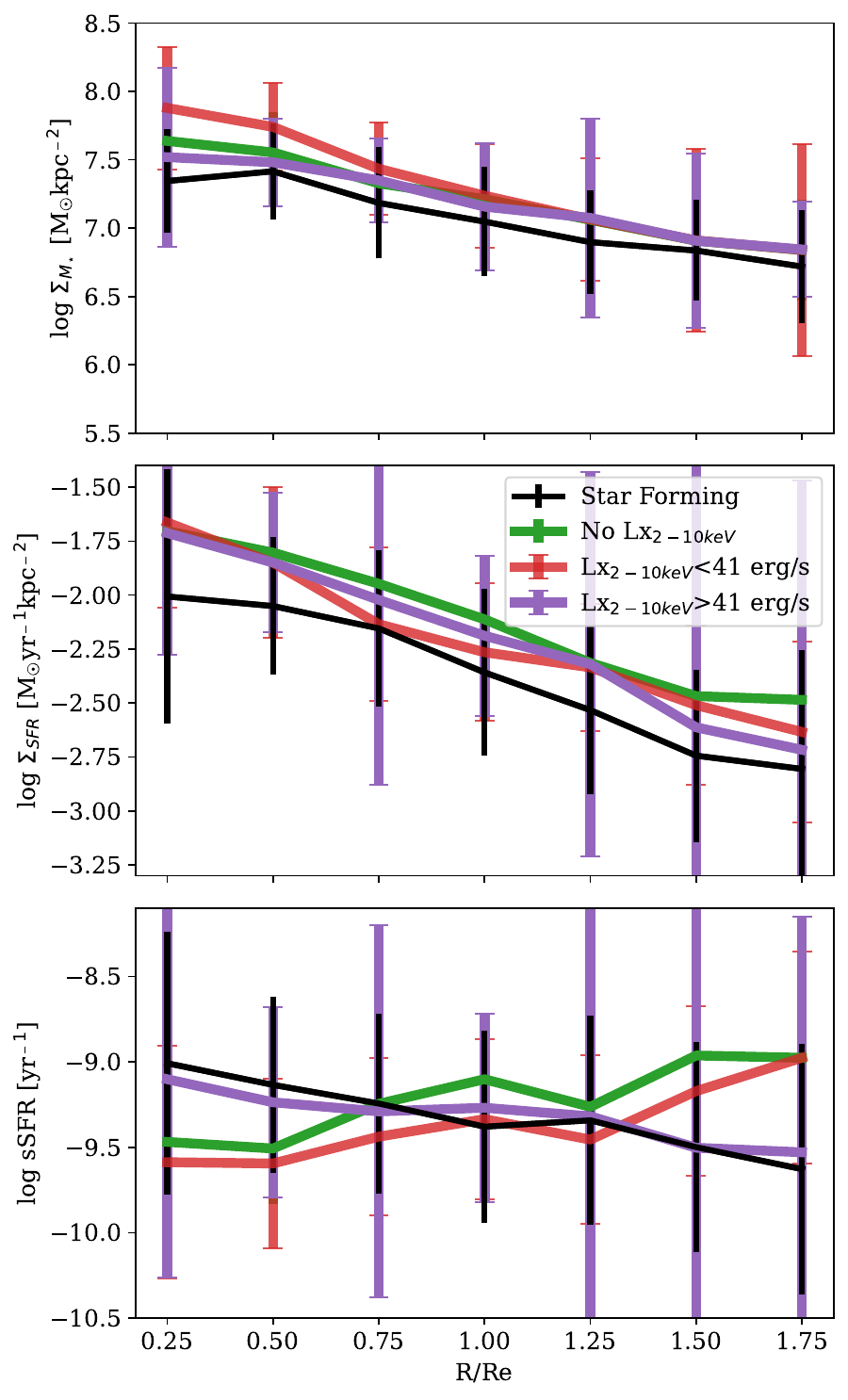}
        \caption{Radial variation in the $\Sigma_{M_\star}$ (top), $\Sigma_{SFR}$ (middle), and sSFR (bottom) of strong and weak AGN based on their X-ray strength compared to the SF sample. The profiles of strong X-ray AGN with log Lx$\mathrm{_{2-10keV}}$ > 41 erg/s are shown with purple lines, moderate X-ray AGN with log Lx$\mathrm{_{2-10keV}}$ < 41 erg/s with red lines, and AGN without Lx$\mathrm{_{2-10keV}}$ detection with green lines. The profiles of SF samples are shown with black lines. This figure shows that strong X-ray AGN exhibit a profile similar to that of the SF sample, whereas weaker X-ray AGN show signs of quenching in the center and enhancement in the outer region of galaxies. Error bars represent the 3$\sigma$ confidence intervals.}
        \label{fig:LxProfilesMassMatched}
    \end{figure}

    To explore if this impact of AGN on host galaxies is an effect of the strength of the X-ray emission, we divided our AGN host galaxies further into weak and strong X-ray emitters and compare the differences between them. We sub-sampled the AGN sample into three bins based on their X-ray strength: AGN without Lx$\mathrm{_{2-10keV}}$ detection (eight sources), 40.1 erg/s < log Lx$\mathrm{_{2-10keV}}$ < 41 erg/s (nine sources), and 41 erg/s > log Lx$\mathrm{_{2-10keV}}$ > 42.3 erg/s (six sources). The radial profiles of AGN and SF sample after matching them in mass are shown in Fig. \ref{fig:LxProfilesMassMatched}. It is clear from the figure that both strong and weak X-ray AGN are found in slightly denser galaxies (by up to 0.5 dex) than the SF galaxies, but are consistent within 3$\sigma$  (Fig. \ref{fig:LxProfilesMassMatched}, top panel). The $\Sigma_{SFR}$ of SF galaxies appear to be lower than all the sub-sampled AGN as well (Fig. \ref{fig:LxProfilesMassMatched}, 2nd panel). We see a complex behavior of AGN host galaxies compared to the SF galaxies when it comes to their sSFR profiles. Their profiles seem to be consistent within 3$\sigma$; however, galaxies with log Lx$\mathrm{_{2-10keV}}$ > 41 erg/s exhibit a very similar sSFR profile to that of the SF sample that is decreasing radially, whereas galaxies undetected in the 2-10keV band  and the ones with log Lx$\mathrm{_{2-10keV}}$ < 41 erg/s appear to be increasing radially (see the bottom panel of Fig. \ref{fig:LxProfilesMassMatched}). In the central regions, the difference between the profiles of weaker and stronger X-ray galaxies seem to stem from a larger mass contained within 0.5R$\mathrm{_e}$ in the weaker X-ray galaxies (see Fig. \ref{fig:LxProfilesMassMatched}, top panel) and the difference in the outer region seem to be driven by the difference in SFR. {The difference in these profiles could be a result of the AGN being in different stages of their duty cycle, which could last from 10$^5$-10$^8$ years \citep{konarParticleAccelerationDynamics2013, schawinskiActiveGalacticNuclei2015, maccagniFlickeringNuclearActivity2020, brienzaRadioSpectralProperties2020} depending on the host galaxy mass \citep{bestHostGalaxiesRadioloud2005, sabaterLoTSSViewRadio2019}. For instance, AGN with the strongest X-ray luminosities might still be in the early stages of AGN evolution and not have had time to suppress star formation, whereas the weaker X-ray AGN, after exhausting their energy, could be at the later stage of their duty cycle and have had enough time to quench their host galaxies.}
    The difference could also simply be due to the very small number of galaxies in each subsample demonstrated by their large error bars. We plan to answer these questions in our upcoming work with a much larger sample of X-ray AGN galaxies detected by the upcoming extended ROentgen Survey with an Imaging Telescope Array (eROSITA) survey \citep{merloniEROSITAScienceBook2012, salvatoEROSITAFinalEquatorialDepth2022} in the J-PAS field. Even though the profiles appear to be consistent within the errors, these profiles reiterate the importance of studying galaxies spatially where we find most interesting and anomalous behaviors.

\subsubsection{Morphological dependence}
\label{sec:morphology}
    
    Morphology is also a factor known to correlate with the star formation and growth of galaxies. For instance, it is known that most late-type galaxies with higher gas fractions exhibit a higher sSFR whereas early types exhibit a lower sSFR due to their lower gas fractions \citep{ealesGalaxyEndSequence2017, caletteHIH_Tostellar2018}. In this section we investigate whether the differences in radial profiles between AGN and control samples are due to morphological differences of the samples. We estimated the bulge-to-disk (B/D) ratio for all the objects in our samples with the estimated magnitude for the bulge and disk obtained from GALFIT using r-band images (see Sect. \ref{sec:removingagn}). B/D ratios quantify the dominance of bulge or disk in galaxies, which give the fraction of stellar mass concentrated in the central bulge compared to the disk. A lower ratio indicates that the object is more likely to be DD, and a higher ratio indicates that the object is more likely to be BD. DD galaxies are generally thought to have a higher SFR than BD galaxies \citep{sersicInfluenceAtmosphericInstrumental1963}. Hence, we divided the control and AGN samples into three bins of B/D ratios and we defined the systems with B/D < 0.5 as DD systems, 0.5 > B/D > 1 as pseudo-bulge (PB) systems, and B/D>1 as BD systems.

\begin{figure}
            \includegraphics[width=0.5\textwidth]{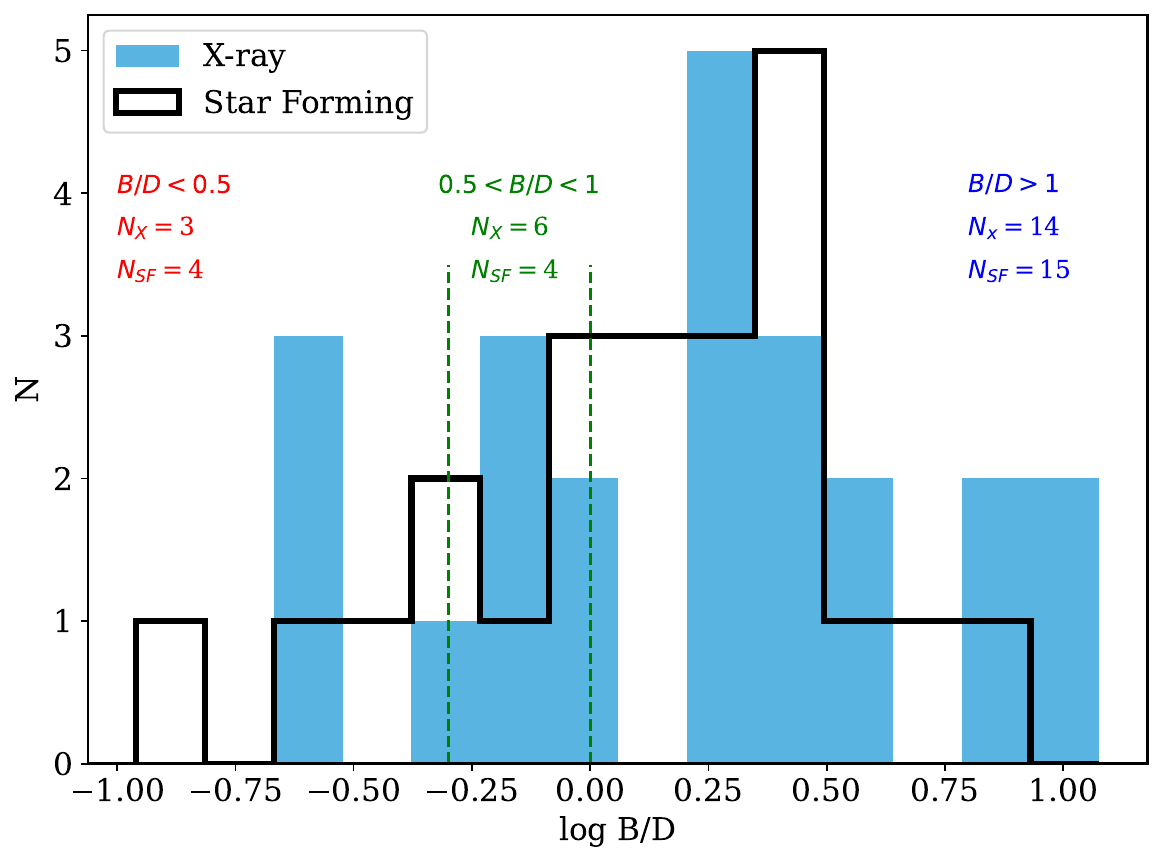}
            \caption{Distribution of log B/D for the AGN and SF samples, shown as blue and black histograms, respectively, for the mass-matched samples. The vertical dashed green lines show the boundaries of  the morphological bins. The numbers of objects in the samples with B/D < 0.5, 0.5 < B/D < 1, and B/D > 1 are displayed in red (left), green (center), and blue (right), respectively.}
            \label{fig:BDratioMassMatched}
        \end{figure}

    We find that 21/32 $\approx 65\%$ of X-ray-selected AGN are hosted in BD galaxies, compared to 41/71 $\approx 58\%$ of SF sample. 5/32 $\approx 16\%$ of the X-ray AGN are hosted in DD galaxies compared to 16/71 $\approx 22\%$ of SF galaxies, while 6/32 $\approx 19\%$ of AGN are hosted in PB galaxies  compared to 14/71 $\approx 20\%$ of SF galaxies. We note here that the results are dominated by low-number statistics, but we find that the findings are consistent with {other studies} that AGN are more likely to be hosted in BD systems than DD systems \citep{kauffmannHostGalaxiesAGN2003, povicAGNhostGalaxyConnection2012,paspaliarisStarformingEarlytypeGalaxies2023}.  However, after matching the samples in mass, we find 3/23 AGN versus 4/23 SF galaxies in DD systems, 6/23 AGN versus 4/23 SF in PB systems and 14/23 versus 15/23 in BD systems in the same mass ranges. The distribution of log B/D values after the samples have been matched in mass is shown in Fig. \ref{fig:BDratioMassMatched}. The results suggest that in a mass-matched samples, there appears to be no significant difference in the morphology of AGN and SF galaxies.
    
\begin{figure*}
            \centering
            \includegraphics[width=17cm]{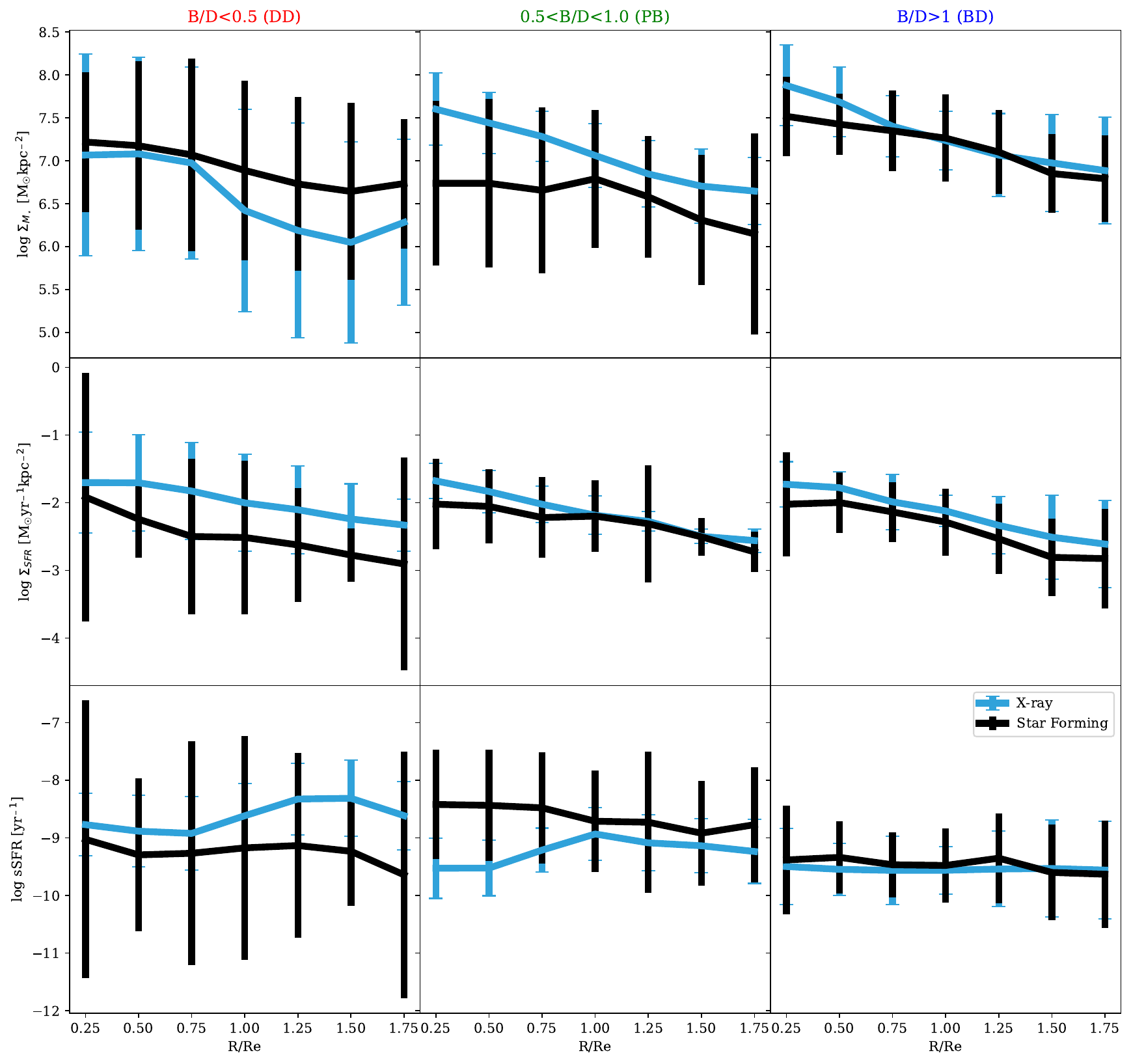}
            \caption{Radial profiles of the $\Sigma_{M_\star}$, $\Sigma_{SFR}$, and sSFR of the AGN and SF samples, sub-sampled on the basis of B/D. The higher the B/D ratio, the more dominant the bulge is in a galaxy. Both samples exhibit decreasing $\Sigma_{M_\star}$ and $\Sigma_{SFR}$ profiles (top two rows). 
            In DD and PB systems, AGN seem to demonstrate a increasing sSFR profile up to 1 $\mathrm{R_e}$ whereas the profiles are flat for both samples in BD galaxies (bottom row).
                Error bars represent the 3$\sigma$ confidence intervals.}
            \label{fig:BDprofilesMassMatched}
        \end{figure*}

    The radial profiles of mass-matched galaxies according to their B/D morphology are presented in Fig. \ref{fig:BDprofilesMassMatched}. It can be seen that in DD systems, AGN and SF galaxies show a similarly decreasing $\Sigma_{M_\star}$ profile up to 0.75 R$\mathrm{_e}$ and are consistent within 3$\sigma$, then the AGN $\Sigma_{M_\star}$ profiles decline very steeply compared to that of SF galaxies. In PB systems however, AGN are seen to be much denser in the central annuli and have a constantly decreasing profile compared to the SF sample, which has a flat $\Sigma_{M_\star}$ profile up to 1 R$\mathrm{_e}$, and then declines as we go toward the edges. In BD systems, both samples show a constantly decreasing $\Sigma_{M_\star}$ profile, and are consistent with each other within 3$\sigma$ (see Fig. \ref{fig:BDprofilesMassMatched}, top row).

    The $\Sigma_{SFR}$ profile of AGN in DD systems is also higher than that of SF galaxies, though within the 3$\sigma$ confidence interval. In PB and BD systems, both samples show a decreasing and consistent $\Sigma_{SFR}$ profile within 3$\sigma$ (see Fig. \ref{fig:BDprofilesMassMatched}, second row). When it comes to their sSFR profiles, it can be seen that AGN and SF galaxies are similar up to 0.75 R$\mathrm{_e}$, then there seems to be an enhancement of star formation in AGN galaxies, whereas the SF galaxies show a flat profile in DD systems (see Fig. \ref{fig:BDprofilesMassMatched}, bottom row). In PB systems, the results are the most interesting where we see that SF galaxies show decreasing sSFR profile, whereas we see the exact opposite trend in AGN galaxies upto 1R$\mathrm{_e}$ and flatten thereafter. In BD systems, both populations show a flat profile where they have a constant sSFR throughout their radial bins, and the profiles are also consistent within 3$\sigma$. The scenarios in DD and PB galaxies showing inside-out quenching could be attributed to the physical processes in the center of AGN host galaxies that release energy in the form of jets, wind, and radiation; this can heat and displace the surrounding gas, preventing its collapse and subsequent star formation. AGN-driven winds and outflows and relativistic jets can interact with the interstellar medium of the host galaxy and inhibit the formation of new stars as well. Regarding the SF properties of SF galaxies, we posit that the $\Sigma_{SFR}$ profiles of our SF sample are in line with \citet{delgadoStarFormationHubble2016} who found a decreasing $\Sigma_{SFR}$ in SF galaxies and that the SF properties are dependent on their morphology and showed that the spheroidal component plays a prominent role in the radial evolution of SF properties of galaxies.

    A caveat here though is that the average masses in DD systems were found to be 10.44 log M$_\odot$ for AGN sample compared to 10.91 log M$_\odot$ for the SF sample. Similarly, in PB systems, the average masses were 10.91 log M$_\odot$ and 10.48 log M$_\odot$ for AGN and SF galaxies. In BD systems, the average masses of two samples were similar with 10.87 log M$\odot$ for the AGN sample and 10.84 log M$_\odot$ for the SF sample, with an average error of $\sim$ 0.6 dex in their masses in all systems. The discrepancy we see in their profiles in the DD and PB systems could stem from the fact that they are not exhibited by galaxies in the similar mass range. Thus, instead of focusing on the absolute quantitative values of the SF parameters, we focused on the trend of radial profiles for these parameters and argue that the AGN sample exhibit an inside-out quenching scenario. 

\section{Discussion}

    \subsection{Magnitude and BPT diagram}
    The filters of miniJPAS provide us with a wealth of information in the optical bands about the intrinsic properties of the galaxies. However, it is also important to note the sample size used in this study. The ideal case scenario for comparing active and inactive galaxies is to match them in redshift, mass, magnitude, luminosity, and size. We were not able to create a control sample that matched in each of these categories due to the challenges posed to us by the very small coverage area of the miniJPAS. In Fig. \ref{fig:SampleProperties} we also see that the r-band magnitudes of AGN host galaxies span to much brighter sources ($\sim$16 mag), compared to the brightest source in our control sample ($\sim$17.5 mag). Although the apparent magnitudes are not the same, the absolute magnitudes are similar, thus assuring us that the comparison in mass is fair, given the direct correlation between absolute magnitude and mass.
    We tried to address this issue by mass-matching the samples as shown in Fig. \ref{fig:MassMatchedResults}. 
    
    In Fig. \ref{fig:BPT}, we can see that two-thirds of our AGN sources lie in the composite region of the BPT diagram, which could also have introduced some bias in the sample. So the results need to be considered with some caution. We used a total of 32 X-ray-selected AGN in the miniJPAS field within $z<0.3$, and only a third of them are in the regime of log Lx$\mathrm{_{2-10keV}}$ > 41 erg/s, with the maximum log Lx$\mathrm{_{2-10keV}}$ of 43.5 erg/s. Two-thirds of the AGN sample had log Lx$\mathrm{_{2-10keV}}$ < 41 erg/s and one-third among them are undetected in 2-10keV band. We also note that 10 of our AGN sample lie in the SF region in the BPT diagram (Fig. \ref{fig:BPT}), which could be a case of optically dull AGN and the subsequent misclassification of X-ray AGN in the BPT diagram \citep{agostinoVLTMUSESpectroscopyAGNs2023}. {To explore the robustness of the  AGN and SF galaxies selection method adopted, we tested our results  using the combination of BPT and {W(Ha) versus [NII]/Ha} (WHAN) diagram \citep{fernandesComprehensiveClassificationGalaxies2011} to select SF and AGN sample. For the SF sample, we selected the mass-matched galaxies that are classified as SF galaxies in both the BPT and WHAN diagrams. Similarly for the AGN sample, we selected X-ray galaxies classified as AGN in both the BPT and WHAN diagrams. This selection method left us with 9 SF and 7 AGN host galaxies. The radial profiles for these subsamples agree with the results found for our fiducial samples,  despite the larger error bars due to lower-number statistics (see Fig. A2).  We plan to explore the selection methods further with a much larger sample in our future work using data from the J-PAS survey.}
    
    \subsection{Large-scale environment and X-ray binaries}
    Studying the impact of the large-scale environment on the radial profiles of galaxies goes beyond the scope of this paper. However, to verify if the contrasting profiles between the two samples is associated with the large-scale environment of the AGN and SF galaxies, we cross-matched our samples against the miniJPAS group catalog by \cite{maturiMiniJPASSurveyCluster2023}. We find that only 4 of the 23 galaxies in the mass-matched SF sample and 6 of the 23 galaxies in the mass-matched AGN sample were linked to groups. In PB systems, where the contrast between the sSFR profiles of the two samples are the most apparent, none of the galaxies were found to be associated with groups or clusters, indicating the absence of a significant role of the large-scale environment for this observation. This finding is consistent with a recent study by \cite{wethersGalaxyMassAssembly2022}, who find that there is no significant difference in the environment of quasars and {inactive} galaxies at 0.1<z<0.35 and that the AGN activity was not dependent on the large-scale environment of galaxies.
    
    Another caveat of this study is the origin of X-rays in our AGN sample. It should be noted that some of the X-ray detections could have originated from LMXBs and HMXBs. We cross-matched the counterparts of X-ray detections with galaxies in the miniJPAS field again within 1" to see if they are emitted from the center and indeed they were, but due to the limitations of observational studies, it is impossible to know their exact location. For example, in an edge-on galaxy, it could be possible that they are produced by X-ray binaries that lie on the outskirts of the galaxy, but we would detect them as central emission due to the viewing angle from our perspective. In principle, LMXBs could prevent the collapse of gas, and suppress SF in their host galaxies, and, if the X-ray emissions did arise from HMXBs, we would expect to see a higher rate of SF activities where they are present, as has been shown in recent observations \citep{soriaChandraVirgoCluster2022} and simulations \citep{vladutescu-zoppDecompositionGalacticXray2023}. The sSFR and $\Sigma_{SFR}$ plots in Fig. \ref{fig:LxProfilesMassMatched} point to the probability of our AGN sample being contaminated by these {HMXBs}. However, past and recent studies of X-ray luminosity functions (see \cite{grimmMilkyWayXrays2002, Fabbiano_2019} for reviews) posit that there is a very low probability that the X-ray emissions (log Lx$\mathrm{_{2-10keV}}$ >39 erg/s) arise from X-ray binaries in the realm of our low redshift AGN sample \citep{lehmerXRayBinaryLuminosity2019, lehmerXRayBinaryLuminosity2020}. We also checked if there were discrepancies in the ages of our strong and weak X-ray AGN derived from the SED fitting that led to the results in Fig. \ref{fig:LxProfilesMassMatched}, which would imply that the galaxies with log Lx$\mathrm{_{2-10keV}}$ < 41 erg/s were still in earlier stages of their evolution, and the stronger ones already evolved in comparison. But we find no such clear relation between their X-ray luminosities and ages (see Fig. \ref{fig:ages}). Studies with many more objects and a higher statistical significance are required to explore the complex relationship between the evolution of X-ray luminosities and spatial star formation activities.
\\
\linebreak
    The limitations of this study were primarily due to our in-ability to create a large, properly matched sample in all the properties as mentioned earlier, due to the small area of miniJPAS. Despite the large errors in radial profiles, this work is a successful proof of concept of the power of the J-PAS photometry as a low resolution IFU. This shows an enormous potential for the studies that J-PAS will enable when combined, for example, with the eROSITA-DE survey (\citealt{merloniEROSITAScienceBook2012, salvatoEROSITAFinalEquatorialDepth2022}) for which there are about 2000 square degrees in common. Besides that, a combination of multiwavelength data in the infrared and radio wavelengths is required to fully understand the outflow of gas and jets that regulate the star formation in AGN hosting galaxies. We plan to address this issue in our next work, which will include a much larger AGN sample, reaching higher redshifts and a greater luminosity range, with a J-PAS catalog of extended sources.\ This will help us better understand and unravel the mysteries surrounding the AGN activity and feedback on their host galaxies.

  \section{Conclusions}
    To unveil empirical evidence of theoretically predicted AGN feedback, we studied the properties of 32 (23 mass-matched) local galaxies with central  X-ray AGN emission at z<0.3, and compared them to a control sample of 71 (23 mass-matched) galaxies at the same redshift within the miniJPAS survey. We extracted stellar mass and SFR estimates from the two samples, taking advantage of the 60 optical band data from miniJPAS after accurate image decompositions via \texttt{GALFITM} to subtract the central AGN component. We acknowledge the low-number statistics of this work and the errors associated with the estimated profiles of galaxies. Thus, {we refrain from making any strong statements, and} rather than focusing on the absolute quantitative values, we focus on the general trend of the radial profiles of our samples. Our main conclusions are as follows.

\begin{itemize}
    \renewcommand\labelitemi{$\bullet$}

    \item AGN host galaxies are generally more massive compared to the inactive population. Even though the mass distributions of  the AGN and SF samples cover a similar range, the median mass for {the AGN sample} is {$\sim$ 0.5 dex} higher than the control sample, though consistent within standard errors (as can be seen in Fig. \ref{fig:GlobalResults}).
    
    \item There is no difference in the total SFR and sSFR of AGN and SF galaxies for the mass-matched samples. This is in line with \citet{mountrichasComparisonStarFormation2022,mountrichasStarFormationXray2022}, {who found no difference between the SF properties of X-ray AGN with log Lx$\mathrm{_{2-10keV}}$ > 42 erg/s and non-AGN galaxies. }We have extended the range down to log Lx$\mathrm{_{2-10keV}}$ $\sim$ 40 erg/s.

    \item The sSFR profiles of the AGN and SF samples are consistent within 3$\sigma$; however, the trend of their radial profiles suggests that AGN quench their host galaxies from the inside out. This is in contrast with the decreasing sSFR profile in SF galaxies, which suggests that the inside-out quenching of galaxies could be connected to the central AGN engine (see Fig. \ref{fig:MassMatchProfile}). These profiles suggest that the reason we do not see a difference in the global SF properties of active and inactive galaxies is due to the simultaneous balancing of suppression (in the center) and enhancement (in the outskirts) of star formation in AGN host galaxies.

    \item When we divide the AGN sample based on hard X-ray luminosity, we find that the AGN 
   undetected in the Lx$\mathrm{_{2-10keV}}$ band and those with log Lx$\mathrm{_{2-10keV}}$ < 41 erg/s exhibit the strongest trends of inside-out quenching, whereas AGN with   log Lx$\mathrm{_{2-10keV}}$ > 41 erg/s show a sSFR profile similar to those of SF galaxies (see Fig. \ref{fig:LxProfilesMassMatched}). {This could be due to the AGN with lower Lx being in their late stage of the AGN life cycle, at which point most of the AGN power has already been exhausted, having already suppressed the star formation in the central regions, while the more powerful AGN could be in their early AGN phase and might not have had enough time to suppress the star formation in the central regions.} We note, however, the large error bars due to the small sizes of the samples. In a future work, combining new, wider, J-PAS data and eROSITA, we plan to study how sSFR profiles depend on nuclear power with significantly higher accuracy.

    \item In DD and PB systems, X-ray AGN show signs of a suppressed center and enhanced outskirts, as inferred from their sSFR profiles. In BD systems, AGN exhibit a flat profile comparable to that of the SF population, and are consistent within 3$\sigma$ confidence intervals (see Fig. \ref{fig:BDprofilesMassMatched}). 
    
\end{itemize}

    \noindent We have studied the difference in star formation parameters between active and inactive galaxies on global to kiloparsec scales and find that, overall, X-ray galaxies have suppressed star formation in the central region compared to their SF counterparts. The findings corroborate the idea that AGN have a limited negative feedback area in their host galaxies that extends up to 1R$\mathrm{_e}$, beyond which the gas might have been blown out {due to AGN feedback}.
    Meanwhile, we have also demonstrated the potential of J-PAS  as a wide-field low-resolution IFU. The advent of J-PAS, which will cover thousands of square degrees of the sky, will allow us to spatially investigate a statistically significant sample of galaxies in the nearby Universe together with their {group- and cluster-}scale environments, which will further our understanding of the coevolution and complex interactions between AGN and their host galaxies.

\begin{acknowledgements}

        This project has received funding from the European Union’s Horizon 2020 research and innovation programme under the Marie Skłodowska-Curie grant agreement No 860744 "Big Data Applications for Black Hole Evolution Sutdies" (BID4BEST). Based on observations made with the JST250 telescope and PathFinder camera for miniJPAS project at the Observatorio Astrofísico de Javalambre, in Teruel, owned, managed and operated by the Centro de Estudios de Física del Cosmos de Aragón. JAFO acknowledges financial support by the Spanish Ministry of Science and Innovation (MCIN/AEI/10.13039/501100011033) and ''ERDF A way of making Europe'' though the grant PID2021-124918NB-C44; MCIN and the European Union -- NextGenerationEU through the Recovery and Resilience Facility project ICTS-MRR-2021-03-CEFCA. S.B. acknowledges support from the Spanish Ministerio de Ciencia e Innovación through project  PID2021-124243NB-C21 and from the Generalitat Valenciana project PROMETEO/2020/085. RGD acknowledges financial support from the Severo Ochoa grant CEX2021-001131-S funded by MCIN/AEI/ 10.13039/501100011033, and PID2019-109067-GB100. RGD, JRM, JMV and GMS acknowledge financial support from the State Agency for Research of the Spanish MCIU through ‘Center of Excellence Severo Ochoa’ award to the Instituto de Astrofísica de Andalucía, CEX2021-001131-S, funded by MCIN/AEI/10.13039/501100011033, and to financial support from the projects PID-2019-109067-GB100 and PID-2022-141755NB-I00. JMV acknowledges financial support from projects PID2019-107408GB-C44 and PID2022-136598NB-C32. IM acknowledges financial support from Severo Ochoa grant CEX2021-001131-S, and PID2019-106027GB-C41 and PID2022-140871NB-C21. AL is partly supported by the PRINMIUR 2017 prot. 20173ML3WW 002 ‘Opening the ALMA window on the cosmic evolution of gas, stars, and massive black holes’ AL is partly supported by the PRINMIUR 2017 prot. 20173ML3WW 002 ‘Opening the ALMA window on the cosmic evolution of gas, stars, and massive black holes. CRA acknowledges support from projects “Feeding and feedback in active galaxies”, with reference PID2019-106027GB-C42, funded by MICINN-AEI/10.13039/501100011033, and ``Tracking active galactic nuclei feedback from parsec to kiloparsec scales'', with reference PID2022-141105NB-I00. JCM acknowledges support from the European Union’s Horizon Europe research and innovation programme (COSMO-LYA, grant agreement 101044612). A. C. acknowledge the financial support provided by FAPERJ grants E-26/200.607 e 210.371/2022(270993). This work made use of Astropy:\footnote{http://www.astropy.org} a community-developed core Python package and an ecosystem of tools and resources for astronomy \citep{astropy:2013, astropy:2018, astropy:2022}.\end{acknowledgements}

\bibliographystyle{aa} 
\bibliography{references}

\appendix

\section{Purity of the AGN and control sample}

\begin{figure}
        \includegraphics[width=0.5\textwidth]{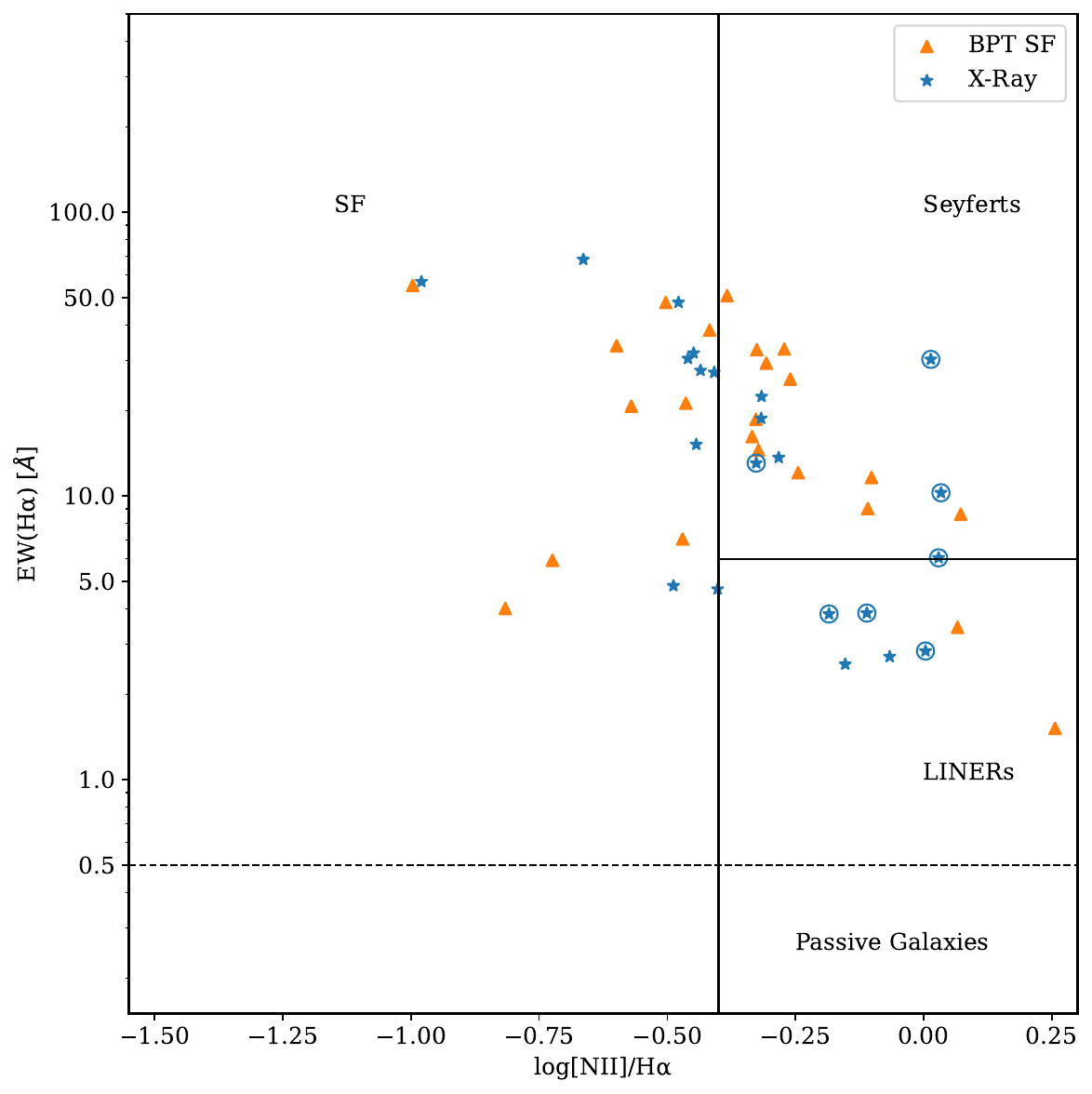}
        \caption{ WHAN diagram used to subdivide our mass-matched SF and AGN samples into various categories. All the objects to the left of -0.6 log[NII]/H${\alpha}$ are SF galaxies, and those below 0.5 H${\alpha}$ in the y-axis are considered to be passive. Galaxies above 0.5 $\AA$ EW(H${\alpha}$) and to the right of  -0.6 log[NII]/H${\alpha}$ are considered to be AGN \citep{fernandesComprehensiveClassificationGalaxies2011}. The blue circles show the X-ray galaxies identified as AGN in both the BPT and WHAN diagrams.}
        \label{fig:WHAN}    
        \end{figure}

\begin{figure}
        \includegraphics[width=0.5\textwidth]{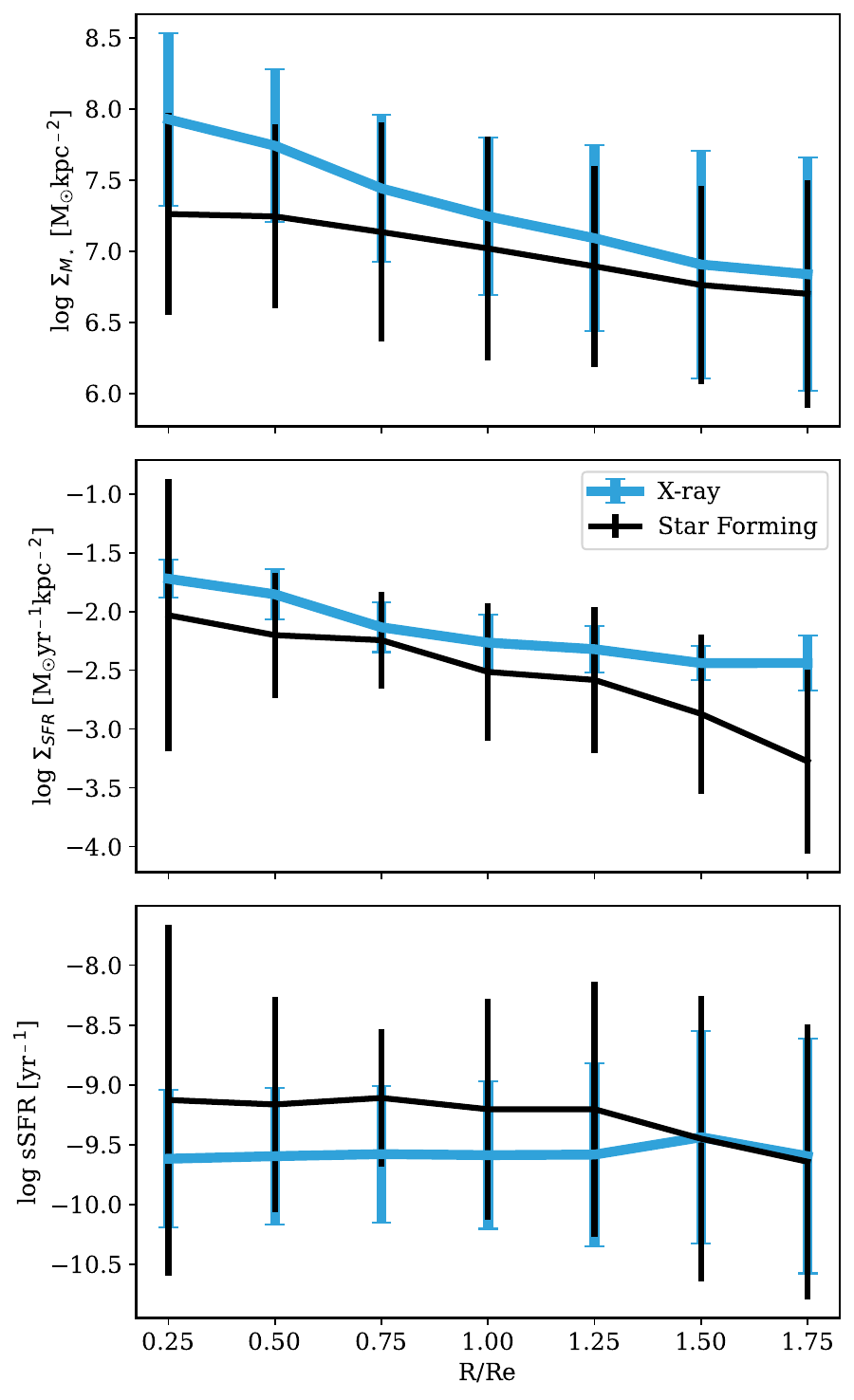}
        \caption{Radial profiles of the $\Sigma_{M_\star}$, $\Sigma_{SFR}$, and sSFR of the AGN and SF samples based on the BPT and WHAN diagrams. We find no significant deviations from the conclusions of the paper. Error bars represent the 3$\sigma$ standard errors.}
        \label{fig:WHANprofile}
    \end{figure}

    To improve the robustness of our work, we further used the WHAN diagram \citep{fernandesComprehensiveClassificationGalaxies2011}. A WHAN diagram is another diagnostic tool similar to the BPT diagram, but it relies only on [NII] and H$\alpha$ lines to distinguish between various classes of galaxies to provide a more robust classification of galaxies with fewer errors. Here, after the mass matching step, we used the WHAN diagram as shown in Fig. \ref{fig:WHAN} to remove Seyferts and Low-ionization nuclear emission-line region (LINER) galaxies from our SF sample and to remove the SF galaxies from AGN sample to see if we get any differences in our results; this left us with nine objects in the SF sample and seven objects in the AGN sample. We did not remove these objects in the main sample selection of this study as it would greatly reduce the number of objects to be analyzed. The possible results are shown here for a general comparison of the results. Figure \ref{fig:WHANprofile} shows the radial profiles of these AGN and SF subsamples. The main conclusion of this paper still applies to these very conservatively selected samples and there is no significant deviation from the main results of this paper; however, we refrain from drawing any strong conclusions as the sample size is too small (cf. Figs. \ref{fig:WHANprofile} and Fig. \ref{fig:MassMatchProfile}).

\section{Validation of the results}

\begin{figure*}
    \includegraphics[width=1\textwidth]{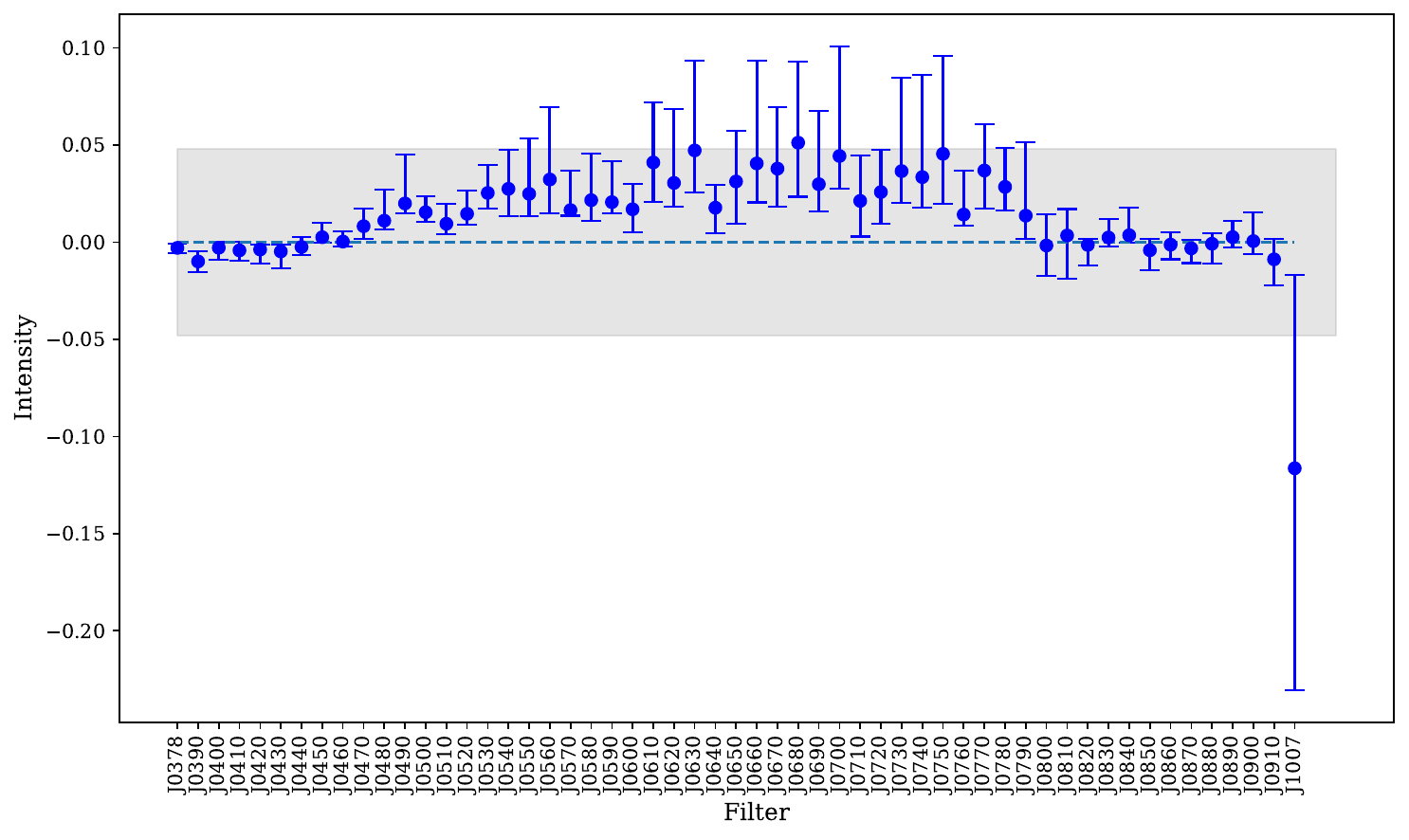}
        \caption{Average intensity of the residuals in the central bins of the AGN sample in each filter obtained after \texttt{GALFITM} modeling. The error bars show the 25th and 75th quantiles of the distribution, and the shaded gray region represents the standard deviation of the mean value. A horizontal dashed line is plotted at intensity = 0, where we would obtain a perfectly flat residual image.}
        \label{fig:AverageResiduals}
\end{figure*}

Figure \ref{fig:AverageResiduals} shows the intensity of average residuals of our X-ray AGN sample obtained from \texttt{GALFITM} modeling. The plot shows that in general, the residuals are close to 0 with an average of 0.015$\pm$0.048 units.

\label{Residuals}

\begin{figure}
    \includegraphics[width=0.5\textwidth]{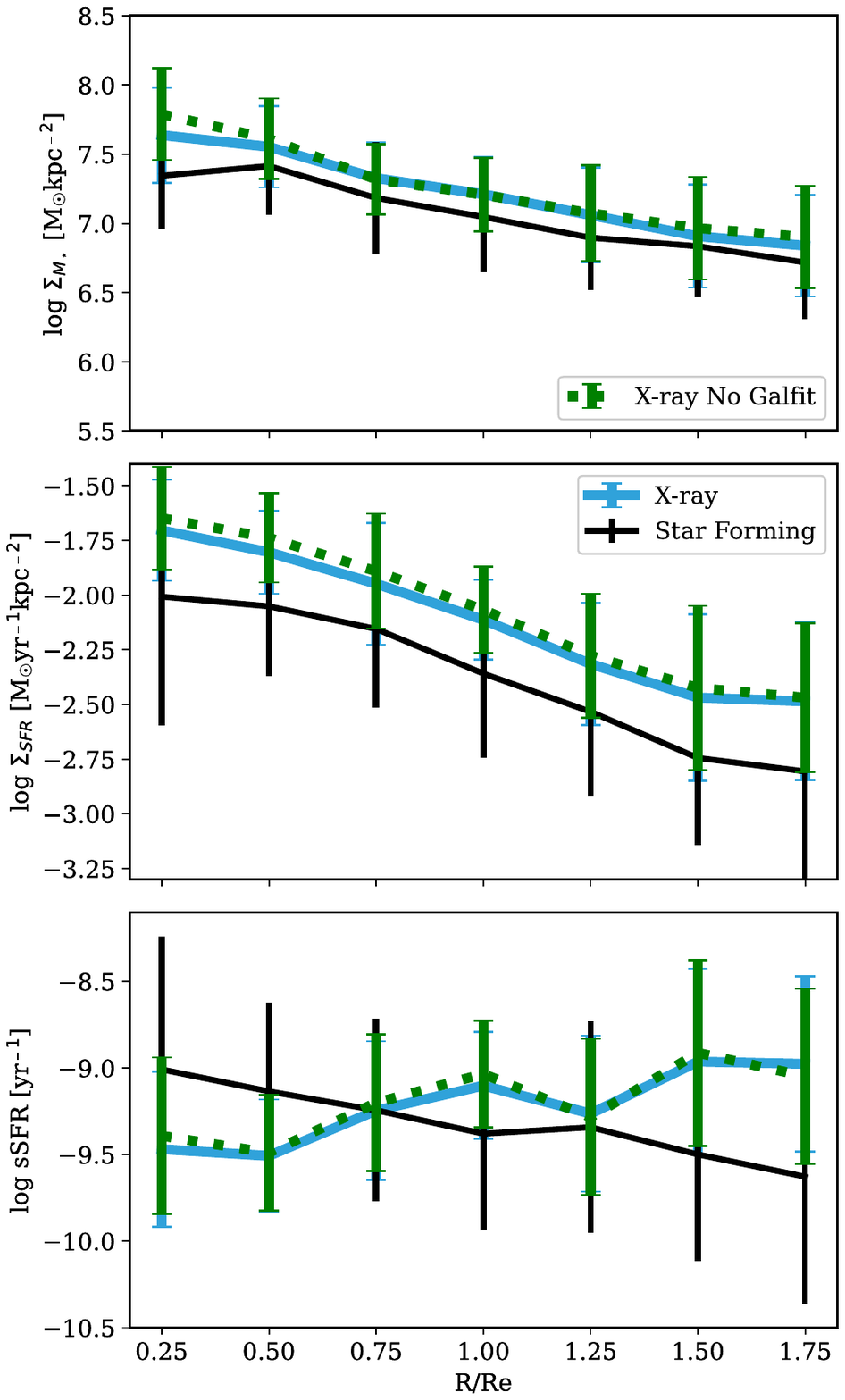}
    \caption{Radial profiles of X-ray AGN and the control sample as shown in Fig. \ref{fig:MassMatchProfile}, with the radial profiles of X-ray galaxies without the AGN subtraction with \texttt{GALFITM} overlaid (dotted green line).}
    \label{fig:GalfitCheck}
\end{figure}

\label{GalfitCheck}
Figure \ref{fig:GalfitCheck} shows the radial profiles of mass-matched X-ray AGN and control sample as shown in Fig. \ref{fig:MassMatchProfile} overplotted by the radial profile of X-ray AGN without the correction of AGN component through \texttt{GALFITM}. The figure suggests that the subtraction of AGN component does not produce any significant difference in the mass and SFR estimations of this work.

\begin{figure}
    \includegraphics[width=0.5\textwidth]{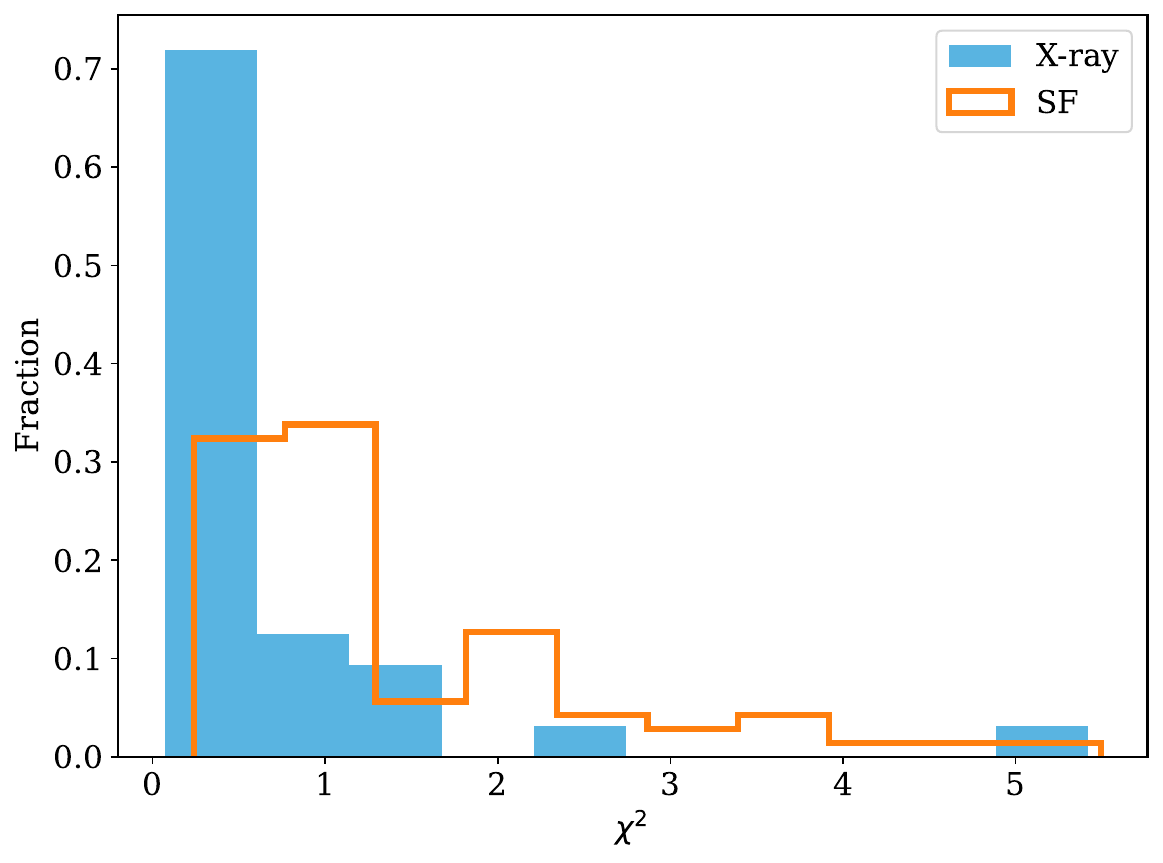}
    \caption{Distribution of reduced $\chi^2$ values of the SED fitting from \texttt{CIGALE}.}
    \label{fig:chisquares}
\end{figure}

\begin{figure}
    \includegraphics[width=0.5\textwidth]{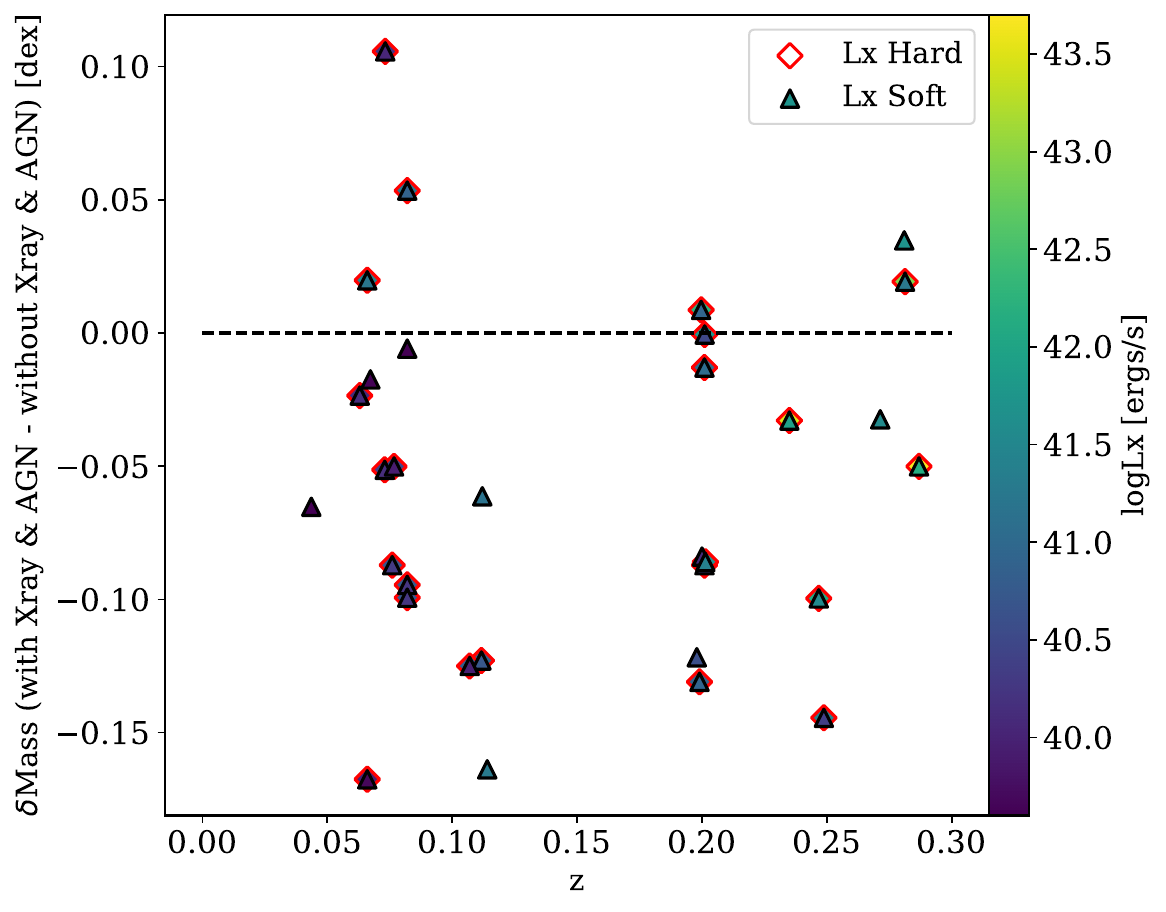}
    \caption{Difference in masses obtained from \texttt{CIGALE} with and without the inclusion of X-ray and AGN modules for the SED fitting of the AGN sample used in this work. The differences in mass for all the objects are within the statistical errors of \texttt{CIGALE}.}
    \label{fig:WithWithoutXrayMass}
\end{figure}

\label{deltaMass}
    {Figure \ref{fig:chisquares}} shows the reduced chi-square distributions of the SED output of \texttt{CIGALE} for our AGN and control sample. Due to the low-number statistics of this work, especially when comparing the number of AGN and SF sample (32 vs. 71), we find the chi-square values of the SED fittings of control sample to be more widely spread throughout the bins compared to the AGN sample.
    We also checked if the inclusion of X-ray and AGN module in \texttt{CIGALE} affects the masses of our AGN sample, and found that the inclusion of these modules constrains the masses by up to 0.16 dex (Fig. \ref{fig:WithWithoutXrayMass}). These values are within the statistical errors of \texttt{CIGALE} output.

\begin{figure}
    \includegraphics[width=0.5\textwidth]{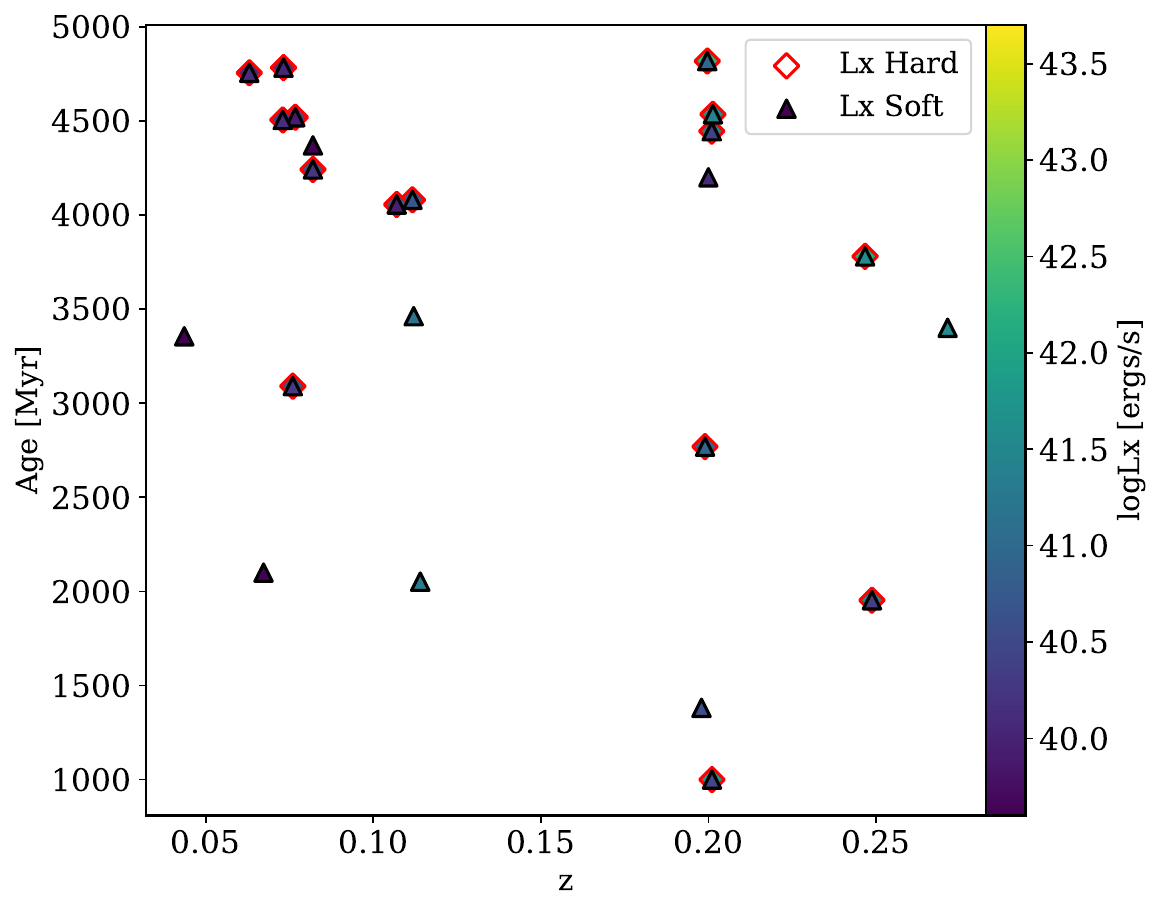}
    \caption{Scatter plot of ages versus redshift, color-coded by X-ray luminosity. No clear correlation is seen between the ages and X-ray luminosities of AGN host galaxies.}
    \label{fig:ages}
\end{figure}

\section{AGN sample}

{In this section we present the list of the AGN host galaxies used in this study in table \ref{tab:AGNsample} and the images along with their SED fitting.}

\begin{table*}[]
\centering
\caption{AGN host galaxies used in this study.}

\begin{tabular}{lllllllll}
\hline
\textbf{JPAS ID} & \textbf{X-ray ID} & \textbf{RA} & \textbf{Dec.} & \textbf{spec\_z} & \textbf{photoz} & \textbf{rMag} & \textbf{\begin{tabular}[c]{@{}l@{}}log Lx (0.2-2 keV)\\ {[}erg/s{]}\end{tabular}} & \textbf{\begin{tabular}[c]{@{}l@{}}log Lx (2-10 keV)\\ {[}erg/s{]}\end{tabular}} \\ \hline
2241-18160       & aegis\_044        & 214.393     & 52.519       & 0.271            & 0.277           & 19.667        & 41.567                                                                            & NaN                                                                              \\
2406-2436        & aegis\_861        & 215.106     & 53.079       & 0.199            & 0.202           & 19.557        & 40.929                                                                            & 40.890                                                                           \\
2406-9782        & egs\_1042         & 215.350     & 53.137       & 0.201            & 0.200           & 19.562        & 40.472                                                                            & 41.776                                                                           \\
2243-7878        & aegis\_553        & 214.772     & 52.883       & 0.076            & 0.075           & 19.509        & 40.317                                                                            & 40.701                                                                           \\
2243-2138        & aegis\_352        & 214.532     & 52.740       & 0.067            & 0.280           & 19.503        & 39.626                                                                            & NaN                                                                              \\
2241-13294       & aegis\_073        & 214.653     & 52.546       & 0.248            & 0.070           & 18.985        & 40.352                                                                            & 41.422                                                                           \\
2241-7071        & egs\_0324         & 214.226     & 52.411       & 0.246            & 0.25            & 19.348        & 41.550                                                                            & 42.067                                                                           \\
2243-3917        & aegis\_262        & 214.778     & 52.670       & 0.197            & 0.197           & 18.995        & 40.595                                                                            & NaN                                                                              \\
2241-7789        & egs\_0336         & 214.268     & 52.415       & 0.281            & 0.25            & 19.505        & 41.246                                                                            & 42.926                                                                           \\
2406-9868        & egs\_1162         & 215.559     & 53.452       & 0.199            & 0.195           & 18.91         & 40.931                                                                            & 42.217                                                                           \\
2470-4691        & egs\_0072         & 213.796     & 51.995       & 0.286            & 0.287           & 18.829        & 42.144                                                                            & 43.654                                                                           \\
2406-5490        & aegis\_888        & 215.301     & 53.106       & 0.201            & 0.202           & 18.571        & 41.394                                                                            & 41.639                                                                           \\
2406-1954        & aegis\_914        & 214.958     & 53.135       & 0.235            & 0.236           & 18.41         & 41.930                                                                            & 43.476                                                                           \\
2243-11625       & aegis\_696        & 215.098     & 52.983       & 0.203            & 0.200           & 18.221        & 40.085                                                                            & NaN                                                                              \\
2243-12120       & aegis\_872        & 214.890     & 53.089       & 0.082            & 0.275           & 17.317        & 40.242                                                                            & 40.707                                                                           \\
2243-8590        & xmmrm\_2272       & 214.542     & 52.986       & 0.114            & 0.112           & 17.645        & 41.288                                                                            & NaN                                                                              \\
2243-8838        & xmmrm\_0875       & 214.517     & 52.987       & 0.112            & 0.114           & 17.772        & 41.164                                                                            & NaN                                                                              \\
2243-6142        & aegis\_601        & 214.861     & 52.921       & 0.082            & 0.200           & 17.449        & 39.615                                                                            & NaN                                                                              \\
2470-14395       & egs\_0138         & 213.941     & 52.224       & 0.111            & 0.197           & 17.032        & 40.716                                                                            & 40.883                                                                           \\
2406-2047        & aegis\_895        & 215.003     & 53.112       & 0.201            & 0.082           & 17.623        & 41.028                                                                            & 41.072                                                                           \\
2243-11362       & aegis\_654        & 215.162     & 52.958       & 0.201            & 0.082           & 17.616        & 40.374                                                                            & 41.376                                                                           \\
2241-16682       & aegis\_008        & 214.595     & 52.452       & 0.281            & 0.104           & 17.825        & 41.719                                                                            & NaN                                                                              \\
2243-9363        & aegis\_541        & 214.672     & 52.871       & 0.107            & 0.114           & 17.062        & 39.961                                                                            & 40.802                                                                           \\
2470-10043       & egs\_0040         & 213.720     & 52.098       & 0.076            & 0.075           & 16.711        & 39.927                                                                            & 40.458                                                                           \\
2243-7944        & aegis\_529        & 214.793     & 52.864       & 0.082            & 0.108           & 16.247        & 40.235                                                                            & 40.782                                                                           \\
2241-10911       & aegis\_296        & 214.529     & 52.697       & 0.066            & 0.074           & 16.431        & 39.676                                                                            & 39.955                                                                           \\
2470-10291       & egs\_0058         & 213.766     & 52.056       & 0.073            & 0.082           & 16.192        & 40.128                                                                            & 40.368                                                                           \\
2470-9821        & egs\_0057         & 213.765     & 52.076       & 0.073            & 0.071           & 16.348        & 40.040                                                                            & 40.206                                                                           \\
2243-14829       & aegis\_819        & 214.770     & 53.047       & 0.082            & 0.059           & 15.27         & 40.887                                                                            & 41.200                                                                           \\
2406-5372        & aegis\_901        & 215.226     & 53.118       & 0.043            & 0.083           & 15.171        & 39.450                                                                            & NaN                                                                              \\
2241-13222       & aegis\_291        & 214.410     & 52.693       & 0.063            & 0.068           & 15.39         & 40.119                                                                            & 40.100                                                                           \\
2241-10941       & aegis\_293        & 214.449     & 52.695       & 0.066            & 0.041           & 15.193        & 41.232                                                                            & 41.400                                                                           \\ \bottomrule

\multicolumn{9}{l}{\footnotesize \textit{JPAS ID}: JPAS ID of the object given by TILE-NUMBER in the miniJPAS database}\\
\multicolumn{9}{l}{\footnotesize \textit{X$-$ray ID}: X-Ray ID of the object obtained from \textit{Chandra} and XMM surveys.}\\
\multicolumn{9}{l}{\footnotesize \textit{RA \& Dec.}: Positions of the sources from the miniJPAS database.}\\
\multicolumn{9}{l}{\footnotesize \textit{specz}: Spectroscopic redshift of the sources obtained from SDSS database.}\\     
\multicolumn{9}{l}{\footnotesize \textit{photoz}: Photometric redshift of the sources obtained from miniJPAS database.}\\     
\multicolumn{9}{l}{\footnotesize \textit{rMag}: Magnitude of sources in the r-band filter from the miniJPAS database.}\\
\multicolumn{9}{l}{\footnotesize \textit{log Lx}: X-ray luminosity of the objects obtained from \textit{Chandra} and XMM surveys.}\\
\multicolumn{9}{l}{\footnotesize {Data used in this work can be accessed from the websites of the respective surveys.}}\\

\end{tabular}
\label{tab:AGNsample}
\end{table*}

\label{XrayThumbs}
    {Figures that show the AGN host galaxies accompanied by their observed J-spectra within their Kron radii and the fitted spectrum with \texttt{CIGALE} to determine their physical properties can be accessed through  Zenodo\footnote{https://zenodo.org/records/11123380}. We note that \texttt{CIGALE} does not fit AGN emission for 1 source with miniJPAS object ID 2241-18160. This object fell in the composite area in the BPT diagram, and in the LINER area in the WHAN diagram. This could represent a case where the AGN is obscured due to dust and \texttt{CIGALE} fails to fit optical AGN emission for this case. This further emphasizes the importance of X-ray emission to identify AGN systems that are optically obscured and are seen as normal galaxies without X-ray detections. We report that the inclusion or emission of this object from our AGN sample does not produce any differences in the results obtained in this paper.

\end{document}